\documentclass[prb,twocolumn,amsmath,amssymb,floatfix,superscriptaddress]{revtex4-1}

\usepackage{hyperref}
\usepackage{graphicx}
\usepackage{tabularx}
\usepackage{bm}
\usepackage{color}
\usepackage[applemac]{inputenc}
\usepackage[applemac]{inputenc}

\newcommand{\beq}{\begin{equation}}
\newcommand{\eeq}{\end{equation}}
\newcommand{\beqa}{\begin{eqnarray}}
\newcommand{\eeqa}{\end{eqnarray}}
\newcommand{\nn}{\nonumber}
\newcommand{\sect}[1]{\emph{#1.---}\ignorespaces}

\newcommand{\editE}[1]{\textcolor{black}{#1}}
\newcommand{\editB}[1]{\textcolor{black}{#1}}
\newcommand{\editP}[1]{\textcolor{black}{#1}}

\begin{document}

\title{\editE{Fluxoid}-induced pairing suppression and near-zero modes in quantum dots coupled to full-shell nanowires}

\author{Samuel D. Escribano}
\author{Alfredo Levy Yeyati}
\affiliation{Departamento de F\'isica Te\'orica de la Materia Condensada, Condensed Matter Physics Center (IFIMAC) and Instituto Nicol\'as Cabrera, Universidad Aut\'onoma de Madrid, E-28049 Madrid, Spain}
\author{Ram\'on Aguado}
\author{Elsa Prada}
\author{Pablo San-Jose}\email[Corresponding author: ]{pablo.sanjose@csic.es}
\affiliation{Instituto de Ciencia de Materiales de Madrid, Consejo Superior de Investigaciones Cient\'{i}ficas (ICMM-CSIC), Madrid, Spain}

\date{\today}
 
\begin{abstract}
We analyze the subgap excitations and phase diagram of a quantum dot (QD) coupled to a semiconducting nanowire fully wrapped by a superconducting (S) shell. We take into account how a Little-Parks (LP) pairing \editE{fluxoid} \editP{(a winding in the S phase around the shell)} influences the proximity effect on the dot. We find that under axially symmetric QD-S coupling, shell \editE{fluxoids} cause the induced pairing to vanish, producing instead a level renormalization that pushes subgap levels closer to zero energy and flattens \editE{fermionic} parity crossings as the coupling strength increases. This \editE{fluxoid}-induced stabilization mechanism has analoges in symmetric S-QD-S Josephson junctions at phase $\pi$, and can naturally lead to patterns of near-zero modes weakly dispersing with parameters in all but the zero-th lobe of the LP spectrum.
\end{abstract}
\maketitle

\sect{Introduction} In the quest to create the necessary conditions for topological superconductivity and Majorana bound states (MBSs)~\cite{Kitaev:PU01,Hasan:RMP10,Qi:RMP11,Alicea:RPP12,Beenakker:RMP15,Sato:ROPIP17,Aguado:RNC17,Prada:NRP20} in hybrid semiconducting nanowires~\cite{Oreg:PRL10,Lutchyn:PRL10,Mourik:S12}, researchers 
have explored new architectures that aim to overcome various limitations in the original nanowire designs. These include shallow two-dimensional quantum wells~\cite{Suominen:PRL17,Fornieri:N19}, nanowires with ferromagnetic coverings~\cite{Liu:AAMI20, Liu:NL20, Vaitiekenas:NP21, Woods:20, Maiani:PRB21, Escribano:PRB21, Liu:arxiv20, Langbehn:PRB21} and full-shell nanowires~\cite{Vaitiekenas:S20, Vaitiekenas:PRB20, Woods:PRB19, Penaranda:PRR20, Valentini:S21, Sabonis:PRL20, Vekris:21,  Kringhoj:PRL21}. The latter, which are the focus of this work, consist of a semiconducting nanowire with epitaxial~\cite{Sestoft:PRM18} superconducting (S) covering on all its facets, instead of only on a few. This apparently simple modification presents some advantages with respect to partial-shell wires and introduces rich new physics. Most notably is the emergence of the Little-Parks (LP) effect under an axial applied magnetic flux~\cite{Little:PRL62,Tinkham:04}. The flux creates \editP{a quantized winding in the phase of the S order-parameter around the shell, also known as \emph{fluxoid}. This} leads to a repeated suppression and re-emergence of the S gap, forming so-called $n = 0,\pm 1,\pm 2\dots$ LP `lobes' as a function of magnetic field, wherein \editP{the fluxoid winding number equals $n$}.

\begin{figure}
   \centering
   \includegraphics[width=\columnwidth]{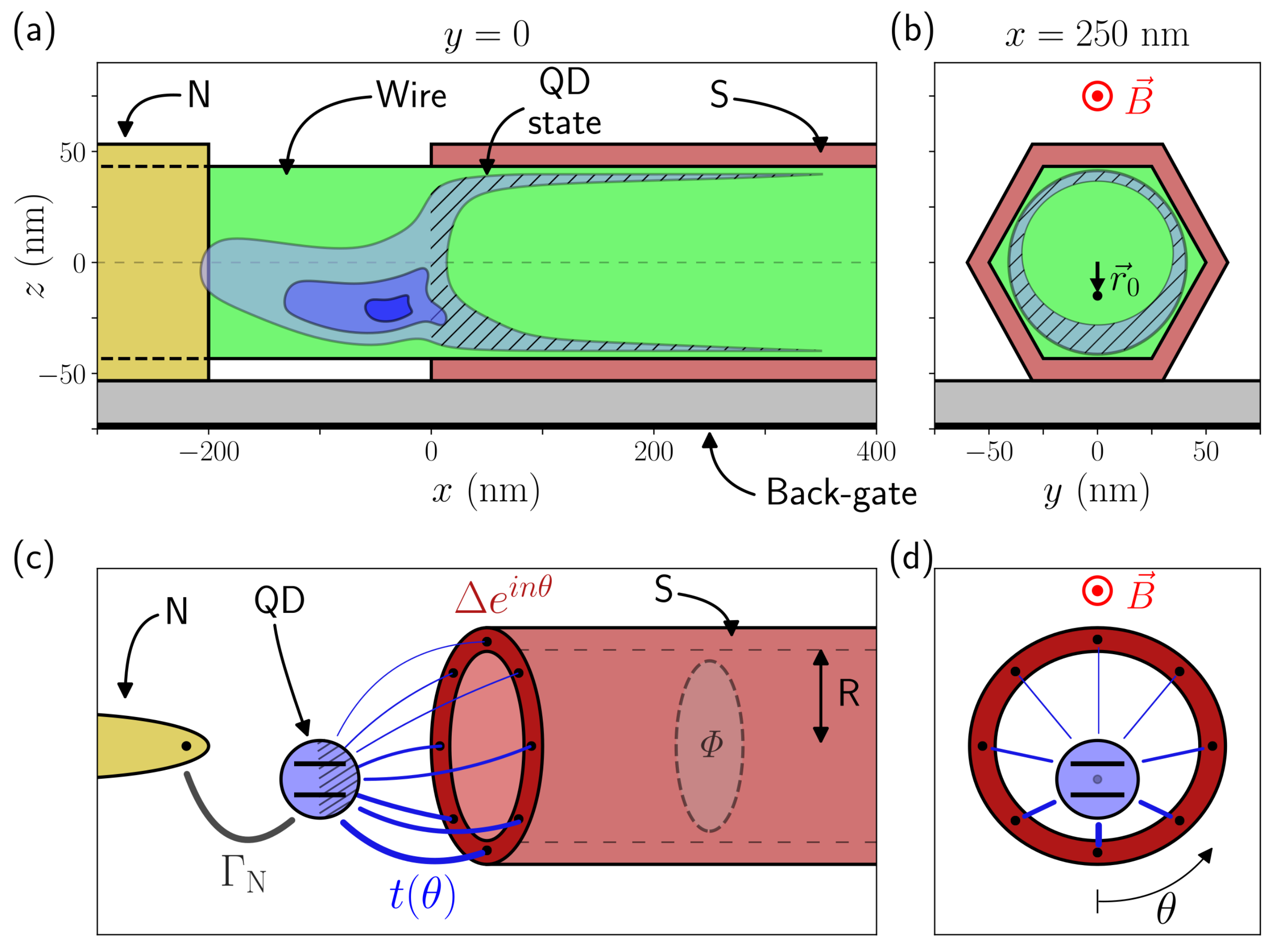}
   \caption{\textbf{QD/full-shell nanowire model.} (a) Schematic lateral cut of a semiconducting  nanowire (green) under a longitudinal magnetic field $B$. The wire is coupled to a normal (N) metal (yellow) to the left and fully wrapped by a superconducting (S) thin shell (red) to the right, with an uncovered segment in between. The junction traps quantum dot (QD) states that can be tuned with a back gate.  A schematic wavefunction of the lowest energy QD state is shown in blue. It is mostly localized in the junction but penetrates into the covered region (dashed blue), where band-bending pushes it towards the superconductor-semiconductor interface. (b) Cross-sectional cut of the same wire inside the full-shell region. $\bm{r}_0=(\langle y\rangle,\langle z\rangle)$ represents the deviation of the wavefunction mean position with respect to wire's axis, informing of its degree of axial symmetry. (c,d) Generalized superconducting impurity Anderson (SIA) model introduced in this work to describe the above hybrid system. A spinful single impurity with energy $\epsilon_0$ and charging energy $U$ is coupled to N with coupling strength $\Gamma_{\rm N}$ and to a hollow S cylinder of radius $R$ with a hopping amplitude $t(\theta)$ that depends on the polar angle $\theta$. A flux $\Phi$ threads the full-shell wire inducing a \editE{winding} in the phase of the S order parameter $\Delta e^{in\theta}$, where the integer $n$ represents the fluxoid quantization and $\Delta$ is the flux-dependent S gap.}
   \label{fig:1}
\end{figure}

Seminal experiments have already demonstrated LP physics in InAs/Al full-shell nanowires~\cite{Vaitiekenas:PRB20}. Concurrently, tantalizing zero-bias anomalies in the tunneling conductance have been observed across $n \neq 0$ LP lobes~\cite{Vaitiekenas:S20}. Explanations in terms of MBSs were put forward~\cite{Vaitiekenas:S20,Penaranda:PRR20}. 
Subsequent experiments reproduced zero bias anomalies in similar devices that were instead explained in terms of quantum dot (QD) Yu-Shiba-Rusinov states~\cite{Yu:APS65,Shiba:PTP68,Rusinov:JL69} localized at the end of the full-shell nanowires~\cite{Valentini:S21}, which are by nature non-topological. \editE{The approximate pinning to zero energy of these states, however, was found to require a specific window of QD-S couplings and a sufficient Zeeman splitting to suppress pairing effects in $n\neq 0$ lobes relative to lobe $n=0$.}

In this work we revisit the problem of a QD coupled to a full-shell nanowire. We study it in the spirit of the superconducting impurity Anderson (SIA) model~\cite{Anderson:PR61}, \editB{extended to explicitly include the cylindrical geometry of the S shell and its pairing \editE{winding} within $n\neq 0$ lobes}, see Fig. \ref{fig:1}, an aspect of these devices not yet analyzed. This \emph{generalized} SIA model allows a previously overlooked mechanism whereby axial \editE{fluxoids} in the shell \editP{suppress pairing effects and} stabilize \editE{fermionic} parity crossings that become increasingly flat versus magnetic field or gates as the QD-S coupling grows. A spontaneous attraction of Andreev bound states (ABSs) towards zero energy develops within all but the $n=0$ LP lobe, where \editP{pairing pushes the ABS} away from zero instead.

The mechanism behind the flattening, or stabilization, of near-zero modes is the cancellation of the proximity-induced pairing potential on the dot states as a result of the pairing phase winding on the shell, combined with a renormalization of the ABS energy due to the strong QD-S coupling. The cancellation requires a sufficient axial symmetry of the dot-nanowire coupling, which is expected owing to the electrostatic screening of the nanowire core by its full-shell Al covering.

\sect{Model} We are interested in tunneling spectroscopy geometries designed to measure the local density of states (LDOS) at one end of the hybrid wire, see Figs. \ref{fig:1}(a,b). Between the full-shell nanowire and the N contact, there is typically an uncovered segment of finite length, required to separate the N tunnel probe from the S shell. The electrostatic potential in the covered semiconductor region is screened from outside electrodes by the shell, and has a dome-like profile in the traverse direction, with a maximum at the nanowire axis, and a minimum at the Al interface due to the band bending induced by the ohmic epitaxial contact~\cite{Sestoft:PRM18, Liu:arxiv20}. On the uncovered region, the potential can instead be controlled by an external back gate. Microscopic simulations show (see Appendix \ref{AppA}) that this geometry naturally leads to the formation of discrete QD-like states, visible in tunneling spectroscopy, that are largely localized in the uncovered segment (see schematic wave function in blue). Their wavefunction also exhibits a long tail seeping into the covered region along the Al interfaces (dashed blue), which allows a strong coupling between the QD state and the epitaxial S shell.


This system can be naturally analyzed within an SIA model~\cite{Valentini:S21}. Its main parameters are the QD charging energy $U$, level energy $\epsilon_0$ (tuned by the back-gate voltage $V_{\rm{bg}}$) and coupling $\Gamma_{\rm{S}}$ to the S contact. This model, however, cannot account for the effect of \editE{fluxoids} on the shell, since it models the S reservoir as a simply connected one-dimensional superconductor. \editB{Here we extend this approach by using instead a hollow,  cylindrical superconductor of radius $R$ and negligible thickness, threaded by an axial flux $\Phi=\pi R^2 B$, where $B$ is the magnetic field applied along the axis, see Figs. \ref{fig:1}(c,d).} This introduces two fundamental differences respect to the conventional SIA model, the emergence of pairing \editE{fluxoids} in the shell and, consequently, a non-trivial proximity effect on the QD due to the nature of the QD-shell coupling, which now involves an integral around the shell perimeter. We discuss their implications below. 

\begin{figure*}
   \centering
   \includegraphics[width=\textwidth]{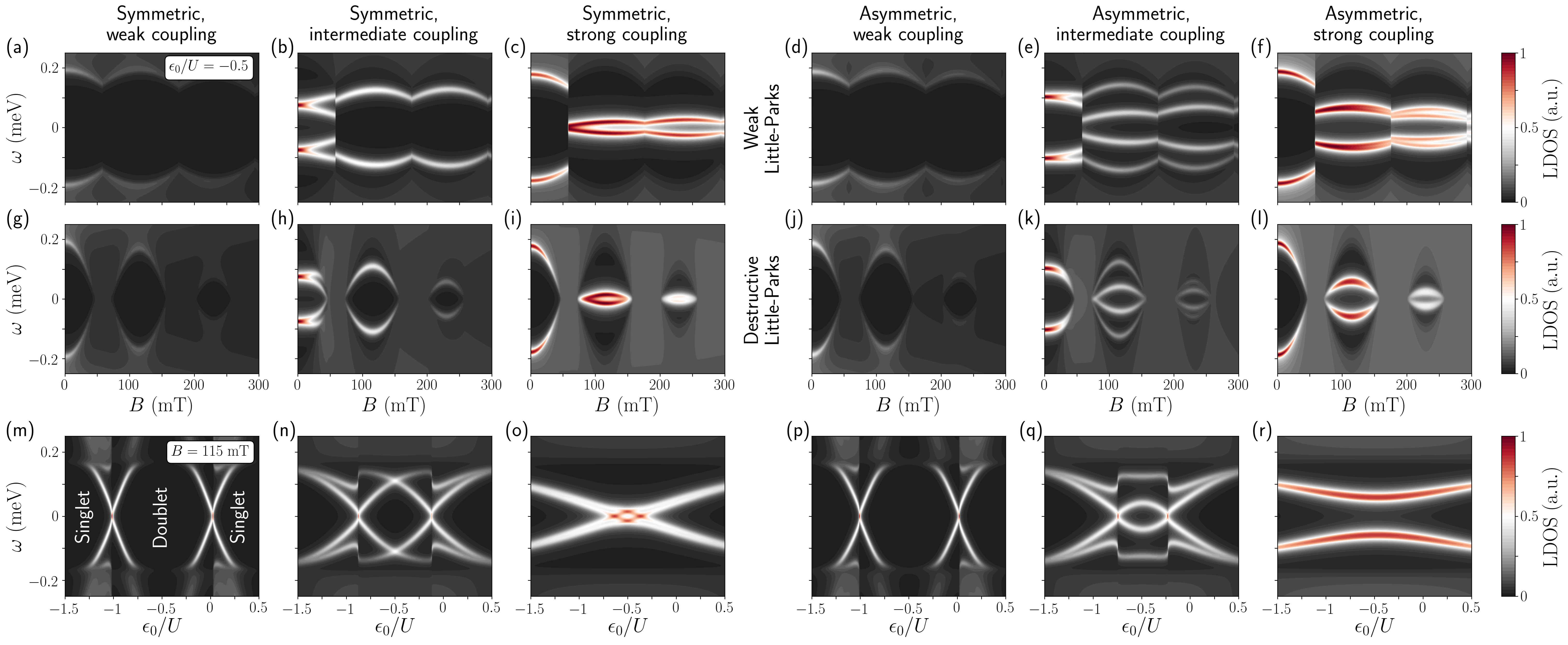}
   \caption{\textbf{QD spectral density.} (Left half) Mean-field local density of states (LDOS) at the QD for a symmetric wave function distribution inside the covered region of the full-shell wire [$\bm{r}_0=0$, see Fig. \ref{fig:1}(b)]. The first, second and third columns correspond, respectively, to the regimes of weak ($\Gamma_{\rm S}/U=0.08$), intermediate ($\Gamma_{\rm S}/U=0.6$) and strong ($\Gamma_{\rm S}/U=2$) coupling between the QD and the superconductor. (Right half) The same but for an asymmetric wave function distribution ($|\bm{r}_0|=R/2$, where $R$ is the shell inner radius). In the first row (a-f) the LDOS is plotted versus the magnetic field $B$ in the weak Little-Parks (LP) regime at half occupation ($\epsilon_0/U=-0.5$), whereas the destructive LP regime is displayed in the second row (g-l). The energy $\omega$ is measured with respect to the superconductor Fermi energy. In the third row (m-r) we show the LDOS versus $\epsilon_0/U$ at finite magnetic field ($B=115$ mT), corresponding to the center of the $n=1$ LP lobe of (g-l). Other parameters: $R=65$ nm, shell thickness $d=25$ nm, InAs g-factor $g=14$, Al gap at zero field $\Delta_0=0.2$ meV, temperature $T=10$ mK, charging energy $U=1$ meV and $\Gamma_{\rm N}=10^{-3}$ meV. 
  The S coherence length is $\xi=185$ nm ($\xi=80$ nm) in the destructive (weak) LP regime.}
   \label{fig:2}
\end{figure*}

The flux into the cylindrical superconductor gives rise to the LP effect, whereby the S gap $\Delta$ becomes modulated by the normalized flux $n_\Phi = \Phi/\Phi_0$ ($\Phi_0=h/2e$ is the S flux quantum), and has maxima at the center of each LP lobe (integer $n_\Phi$). A Ginzburg-Landau treatment can be used to compute the precise $\Delta(n_\Phi)$ dependence \cite{Larkin:05,Liu:S01,Schwiete:PRL09,Schwiete:PRB10}. Depending on the coherence length $\xi$, this includes a \emph{destructive} LP regime ($R/\xi < 0.6$) where $\Delta$ vanishes between lobes, or a \emph{weak} LP regime where $\Delta$ remains finite ($R/\xi\gtrsim 1$).
The complex S pairing $\bm\Delta$ furthermore develops a \editE{winding} $n$ within the $n$-th lobe except $n=0$, $\bm\Delta(\theta)=\Delta(n_\Phi)\exp\left(i n\theta\right)$, where $\theta$ is the polar angle around the shell and $n = \left \lfloor{n_\Phi}\right \rceil$.

In the spirit of the SIA model, the QD is modeled as a point at position $\bm{r}_0$. The QD is coupled to points $\bm{r}_\theta = (0, R\cos\theta, R\sin\theta)$ on the inside \footnote{Note that the QD-S coupling in the system of Figs. \ref{fig:1}(a,b) is given by the overlap between the wavefunction inside the shell (shaded region) and the superconductor. The wavefunction in the uncovered region will instead mainly influence the level position $\epsilon_0$.} of the shell by a hopping amplitude $t(\theta)$, see Figs. \ref{fig:1}(b,c). In an axially symmetric situation (on-axis QD, $\bm r_0=0$), the hopping $t(\theta)=t_0$ is a constant. The axial symmetry may be broken by considering a displacement of $\bm{r}_0$ away from the nanowire axis, microscopically representing a deviation from axial symmetry of the dashed region of the QD state wave function of Figs. \ref{fig:1}(b), and thus a non-constant $t(\theta)$ (see the Appendix \ref{AppB} for further details on QD modeling).


The extended SIA model Hamiltonian takes the form
\beqa
H &=& H_{\rm D} + H_{\rm S} + V_{\rm SD},\\
H_{\rm D} &=& \sum_\sigma(\epsilon_0 + \sigma V_{\rm Z})d^\dagger_\sigma d_\sigma + U n_\uparrow n_\downarrow, \nn\\
H_{\rm S} &=& \int dz\,d\theta\sum_\sigma\left[\psi^\dagger_{\sigma\theta x}\frac{\bm p^2}{2m^*}\psi_{\sigma\theta x} \right.\nn\\
&&\left.+ \Delta(n_\Phi) e^{in\theta}\psi^\dagger_{\sigma\theta x}\psi^\dagger_{-\sigma\theta x} + \mathrm{h.c}\right], \nn\\
V_{\rm SD} &=& \int d\theta\sum_\sigma t(\theta) \psi^\dagger_{\sigma\theta 0}d_\sigma + \mathrm{h.c}.\nn
\eeqa
In the above equations, QD operators of spin $\sigma$ are denoted by $d_\sigma$ and shell operators by $\psi_{\sigma\theta x}$, where $x$ is the coordinate along the shell ($x=0$ denotes its rim). The QD occupation operators are $n_\sigma = d^\dagger_\sigma d_\sigma$, $U$ is the charging energy and $V_{\rm Z}=\frac{1}{2}g\mu_B B$ is a Zeeman field (with Land\'e g-factor $g$ and Bohr magneton $\mu_B$). \editE{The $\theta$ dependence of the pairing phase and $t(\theta)$ is the key difference between our model and the conventional SIA model used in Ref. \onlinecite{Valentini:S21}.}

The proximity effect on the QD is apparent upon integrating out the shell, which yields a Green's function $G^{\rm D}(\omega) = \left[\omega - H^{\rm D}_\mathrm{\rm eff}(\omega)\right]^{-1}$ in terms of an effective Hamiltonian $H^{\rm D}_\mathrm{\rm eff}(\omega)$ that includes self-energies from the S and also from QD interactions $U$. For an axially symmetric QD at $\bm r_0=0$ we obtain 
\beqa
\label{Heff}
H^{\rm D}_\mathrm{\rm eff}(\omega) &=&H^\mathrm{\rm D}_{\rm eff,\uparrow}(\omega)\oplus H^\mathrm{\rm D}_{\rm eff,\downarrow}(\omega),\nonumber\\
H^\mathrm{\rm D}_{\rm eff,\sigma}(\omega) &=& \left(\begin{array}{cc}
\epsilon_0 + \sigma V_Z & 0\\
0 & -\epsilon_0 + \sigma V_Z
\end{array}\right)+\Sigma^{\rm U}_\sigma +\Sigma^{\rm S},\nn\\
\Sigma^{\rm U}_\sigma &\approx& U
\left(\begin{array}{cc}
\langle n_{-\sigma}\rangle & \langle d_{\sigma}d_{-\sigma}\rangle\\
\langle d^\dagger_{\sigma}d^\dagger_{-\sigma}\rangle  & - \langle n_{\sigma}\rangle
\end{array}\right),\nn\\
\Sigma^{\rm S} &\approx& -\frac{\Gamma_{\rm S}}{\sqrt{\Delta^2-\omega^2}}
\left(\begin{array}{cc}
\omega & \Delta \delta_{n,0}\\
\Delta \delta_{n,0} & \omega
\end{array}\right),
\label{SigmaS}
\eeqa
where $\Gamma_{\rm S} \equiv \pi t_0^2\rho_{\rm S}$ and $\rho_{\rm S}$ is the density of state of normal Al at the Fermi energy.  An additional, small self-energy $\Sigma^{\rm N}=-i\Gamma_{\rm N}$ from the N tunnel probe is also present, but is omitted above for simplicity. $\Sigma^{\rm U}_\sigma$ includes interactions at a self-consistent mean-field level (\editP{Kondo correlation effects are discussed in the Appendix \ref{AppC}}).
\editB{Crucially, the off-diagonal pairing term in $\Sigma^{\rm S}$ depends on the lobe index $n$ through the Kronecker delta  $\delta_{n,0}$, which results in a suppressed S pairing on the QD in all lobes except $n=0$. This has to be contrasted with the standard SIA model} \editE{as discussed in Ref. \onlinecite{Valentini:S21}}, \editB{where the Kronecker delta is absent,} \editP{and a strong Zeeman field was instead invoked within $n\neq 0$ lobes to reduce pairing effects.
The suppression of pairing in our model} is robust even with interactions, since a positive $U$ will not introduce pairing corrections unless $\langle d_{\sigma}d_{-\sigma}\rangle\neq 0$ to start with. Breaking axial symmetry ($\bm r_0\neq 0$) gradually restores pairing at all lobes (see the Appendix \ref{AppB} for the general form of $\Sigma^{\rm S}$.)

To understand the observable consequences of Eq. \eqref{SigmaS} it is worthwhile to take a moment to discuss its physical origin. Hoppings from the QD to a point around $\bm{r}_\theta$ in the shell and back produce a contribution to $\Sigma^{\rm S} = \int d\theta\Sigma^{\rm S}(\theta)$ of the form $\Sigma^{\rm S}(\theta) = 
\int d\bar\theta\,t(\theta+\bar\theta) g^{\rm S}(\theta+\bar\theta, \theta-\bar\theta) t(\theta-\bar\theta)$, where $g^{\rm S}(\theta,\theta')$ is the Green's function of the decoupled superconductor between two points $\bm{r}_\theta, \bm{r}_{\theta'}$ around the rim. For a symmetrically coupled QD, this $\Sigma^{\rm S}(\theta)$ can be easily shown to have an off-diagonal pairing term proportional to $\bm{\Delta}(\theta)=\Delta e^{in\theta}$ itself, which is zero when integrated over $\theta$, unless $n=0$. In other words, the off-diagonal $\Sigma^{\rm S}(\theta)$ inherits the phase of the \editE{fluxoid} at $\bm{r}_\theta$, and thus cancels out for a constant $t(\theta)=t_0$ upon integrating $\theta$. The diagonal part of $\Sigma^{\rm S}$, however, does not suffer this cancellation, and is always present. Its effect, in the absence of an off-diagonal pairing term, is to renormalize the state energy, pushing it closer to zero. Indeed, assume that the electron part of the QD Green's function $G^{\rm D}_{11}(\omega)\sim(\omega-\Omega_0)^{-1}$ has a pole at $\omega=\Omega_0$ representing a dot level at $\epsilon = \mathrm{Re}(\Omega_0)$ when decoupled from the superconductor (i.e. for $\Sigma^{\rm S}=0$ but finite $\Gamma_{\rm N}$ and $U$). The coupling to the S shell induces a purely diagonal $\Sigma^{\rm S}(\omega)$ within any $n\neq 0$ lobe that modifies the pole following
\beq
\Omega-\Omega_0 - \Sigma^{\rm S}_{11}(\Omega) = \Omega-\Omega_0 +\frac{\Gamma_S\Omega}{\sqrt{\Delta^2-\Omega^2}}=0,
\eeq
which, for $\epsilon\ll\Delta$, yields a renormalized level position
\beq
\label{renorm}
\epsilon' = \mathrm{Re}(\Omega) = (1+\Gamma_S/\Delta)^{-1} \epsilon.
\eeq
Any QD state deep inside the gap will thus see its energy scaled down by a factor $(1+\Gamma_{\rm S}/\Delta)^{-1}$, shifting it ever closer to zero as we increase the coupling $\Gamma_{\rm S}$. This is exactly the opposite of the effect of $\Sigma^{\rm S}$ for $n=0$, which tends to push the state towards the gap edge by virtue of the off-diagonal pairing term \footnote{This simple argument assumes no $\Gamma_{\rm S}$-dependent corrections to $\Sigma_{\rm U}$, but it is qualitatively confirmed by our mean field calculations.}.

\sect{Results} We now present numerical simulations of the LDOS in the QD, $\rho(\omega) = -\frac{1}{\pi}\rm{Im}\,\rm{Tr}\,G^{\rm D}(\omega)$, as a function of magnetic field $B$, bare QD level $\epsilon_0$, coupling $\Gamma_{\rm S}$ and coupling asymmetry $|\bm{r}_0|/R$. Figure \ref{fig:2} summarizes the results, see caption for details. 

The symmetric $\bm{r}_0=0$ case is displayed in the first three columns for increasing coupling $\Gamma_{\rm S}/U$ 
For strong coupling, Figs. \ref{fig:2}(c,i), an ABS near the gap edge at $n=0$ transforms into a stable, non-topological, near-zero-energy anomaly across all $n\neq 0$ lobes. This resonance is composed of four near-zero peaks, instead of the two peaks expected from a (split) Majorana zero-bias anomaly. 

\editP{In the third row, the LDOS is calculated against $\epsilon_0/U$ at the center of the $n=1$ destructive LP lobe. Unlike in the standard SIA model, the zero-energy parity crossings of the lowest excitation in the symmetric case are not destroyed by a strong coupling, nor is the crossing at $\epsilon_0/U=0.5$ with the second excited state. Instead, the excitations evolve into an increasingly flat double-X pattern, Fig. \ref{fig:2}(o). As discussed above, this is a result of the vanishing pairing, Eq. \eqref{SigmaS}. The asymmetric case in the last three columns recovers a more standard SIA behavior, with no flattened zero modes at strong coupling.}

\begin{figure}
   \centering
   \includegraphics[width=\columnwidth]{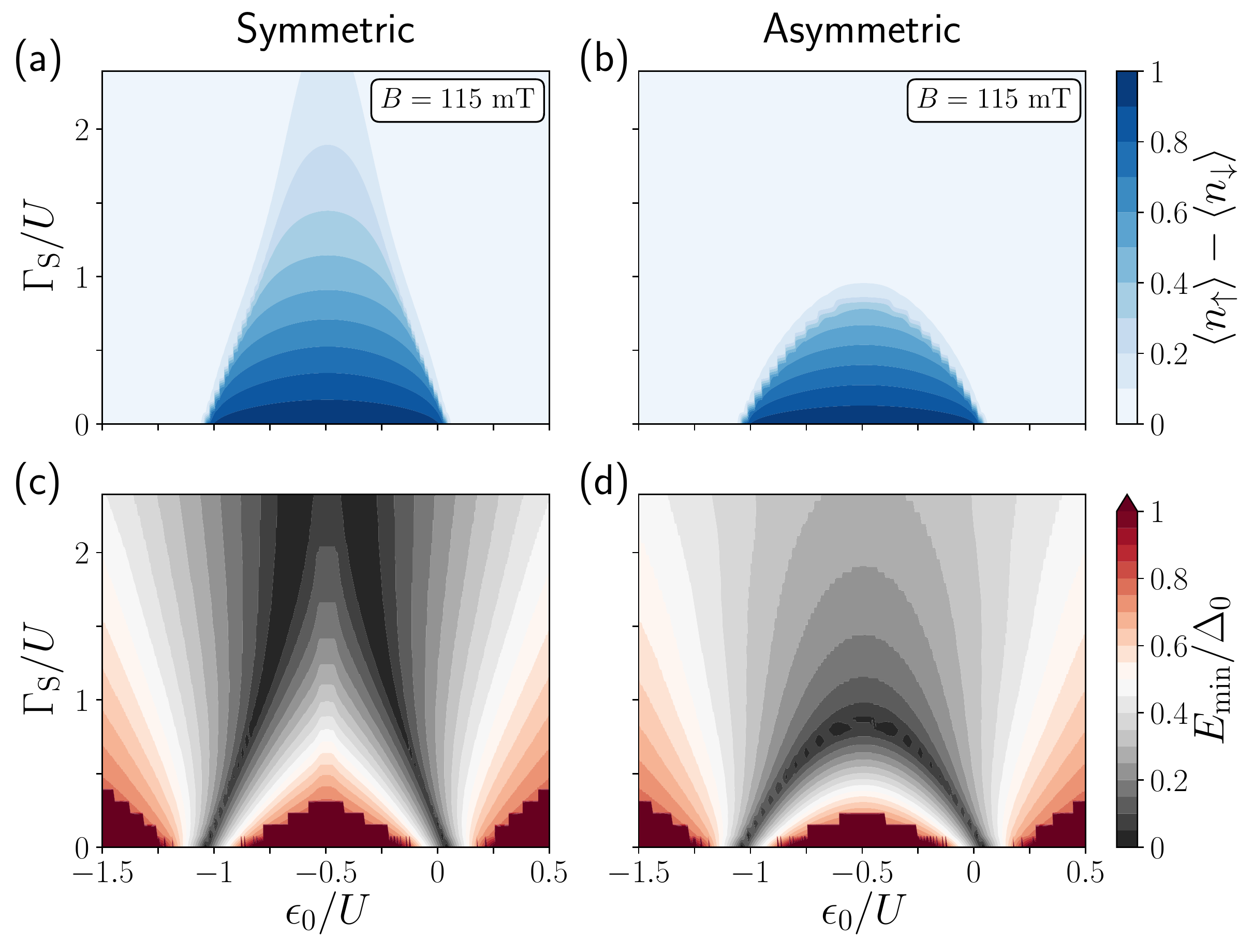}
   \caption{\textbf{QD phase diagram.} (a,b) Spin polarization  $P = \langle n_{\uparrow}\rangle-\langle n_{\downarrow}\rangle$ of the QD level vs $\Gamma_{\rm S}/U$ and $\epsilon_0/U$ at the center of the $n=1$ destructive LP lobe. In the case (a) of a symmetric coupling ($\bm{r}_0=0$), the singlet-doublet phase transition boundary displays a ``chimney"-like shape where the doublet ground state phase ($P\neq 0$) extends to arbitrarily large values of $\Gamma_{\rm S}/U$. In the asymmetric case (b) ($|\bm{r}_0|=R/2$ here) we recover a ``dome"-like shape for the singlet-doublet boundary, typical of the standard SIA, above which the ground state is a singlet ($P=0$). (c,d) Minimum excitation energy $E_\mathrm{min}$ normalized to the zero-field Al gap $\Delta_0=\Delta(0)$. The singlet-doublet boundary is associated to $E_\mathrm{min} = 0$ \editE{fermionic} parity crossings (black). Rest of parameters as in Fig. \ref{fig:2}.}
   \label{fig:3}
\end{figure}

The above phenomenology can be collected into a phase diagram in a $\Gamma_{\rm S},\epsilon_0$ phase space, in terms of the ground-state spin-polarization $P = \langle n_\uparrow\rangle - \langle n_\downarrow\rangle$ (zero for singlet, finite for doublet ground states) or in terms of the energy $E_\textrm{min}$ of the lowest excitation. This is shown in Fig. \ref{fig:3} for the symmetric and asymmetric cases. The latter (b,d) shows a dome like doublet phase as in the \editP{standard} SIA model. In contrast, the symmetric case (a,c) shows a ``chimney"-like doublet phase extending all the way to $\Gamma_{\rm S}\to\infty$, with an $E_\textrm{min} \approx \frac{1}{2}E_c/(1+\Gamma_{\rm S}/\Delta)$ at $\epsilon_0 = -U/2$ according to Eq. \eqref{renorm}, where $E_c\propto U+2V_Z$ is the asymptotic width of the chimney region at large $\Gamma_{\rm S}/U$.

The symmetric $\bm{r}_0=0$ phase diagram and LDOS are reminiscent of the one in symmetric S-QD-S junctions at fixed phase difference $\phi=\pi$ ~\cite{Vecino:PRB03,Oguri:JPSJ04,Zonda:SR15,Kirsanskas:PRB15,Martin-Rodero:AIP11}. 
Interestingly, such a system exhibits the same structure of $\Sigma^{\rm S}$ in the QD, with a cancellation of the induced pairing from the two symmetrically-coupled S terminals due to their $\pi$ phase difference, and hence a similar chimney-like singlet-doublet phase boundary. 

\sect{Conclusions} We have introduced a generalized SIA model to understand the phase diagram of a QD coupled to a cylindrical S shell. The combination of axial LP \editE{fluxoids} in the shell and a highly symmetric QD-S coupling produces an anomalous type of proximity effect, similar to that of $\pi$-phase QD-Josephson junctions. Microscopic simulations show that a symmetric QD-S coupling is in fact the natural physical situation in these devices that requires no particular tuning, since it is controlled by the part of the QD wavefunction in contact with the shell, which is screened from the external electrostatic environment (see Appendix \ref{AppA}). \editE{Only defective or disordered full-shell wires, or wires with a noticeably asymmetric cross-section would produce a non-negligible QD-S coupling asymmetry.}
At strong \editP{and symmetric} QD-S coupling, the anomalous, \editE{fluxoid}-induced proximity effect stabilizes near-zero states within $n \neq 0$ LP lobes, while conventional proximity pushes them away from zero at the $n=0$ lobe. This kind of ``antipairing'' effect for $n\neq 0$ is a strong departure from the phenomenology of conventional SIA models, and predicts a subgap LP spectral pattern \editE{that would result in near-zero bias peaks in local spectroscopy measurements in full-shell nanowires. 
The effect of this mechanism on the length dependence of Coulomb blockade period modulation, which was reported to be exponentially decaying with island length in Ref. \onlinecite{Vaitiekenas:S20}, remains an important open question.}

\acknowledgments
This research was supported by the Spanish Ministry of Economy and Competitiveness through Grants FIS2017-84860-R, PCI2018-093026 and PGC2018-097018-B-I00 (AEI/FEDER, EU), the European Union's Horizon 2020 research and innovation programme under the FETOPEN grant agreement No. 828948 (AndQC) and the Mar\'ia de Maeztu Programme for Units of Excellence in R\&D, Grant No. MDM-2014-0377.

\appendix

\section{Microscopic model} 
\label{AppA}

In this section, we analyze the electronic structure and the three-dimensional (3D) electrostatic screening (including the surrounding media) of an InAs nanowire of length $L$, with a right portion of length $L_S$ covered all around by a thin superconducting (S) Al shell, contacted on the left to a normal (N) electrode, and with an uncovered junction region of length $L-L_S$ in between, see Figs. \ref{fig:1}(a,b) of the main text. A back gate extends below the whole wire, but is only effective to tune the electrostatic potential in the uncovered region. This particular geometry has been explored experimentally in the context of Majorana bound states~\cite{Vaitiekenas:S20, Valentini:S21}. 

The purpose of this section is to show that this structure develops low-energy states, which we call quantum-dot (QD) states, that are mostly localized in the uncovered region but that extend well into the full-shell nanowire, where they tend to be close to the superconductor-semiconductor interface. We will show that the wavefunction profile inside the covered region is essentially axially symmetric independently of the backgate voltage, which is screened inside the shell portion.


The Hamiltonian of the system, sketched in Fig. \ref{fig:SM0}(d), is defined as
\begin{eqnarray}
H_{\rm 3D}=\left(\frac{\hbar^2k^2}{2m^*}-e\phi(\bm{r})\right)\sigma_0\tau_z + \Delta(\bm{r})\sigma_y\tau_y \nonumber \\
 + \frac{1}{2}\left[\bm{\alpha}(\bm{r})\cdot \left(\bm{\sigma}\times\bm{k}\right)+\left(\bm{\sigma}\times\bm{k}\right)\cdot \bm{\alpha}(\bm{r})\right],
\label{Eq:H_3D}
\end{eqnarray}
where $-e$ is the electron charge, $m^*$ is the semiconductor effective mass, $\phi(\bm{r})$ the electrostatic potential inside the wire, $\Delta(\bm{r})$ is the induced superconducting pairing, and $\bm{\alpha}(\bm{r})$ the spin-orbit coupling (SOC). We perform this study at zero magnetic field $B=0$ as a finite field of a few hundred mT does not significantly influence the wavefunction profile. Notice that this Hamiltonian only applies to the semiconducting nanowire across the two areas of interest: the uncovered and covered regions. For numerical efficiency reasons we do not include in the Hamiltonian a microscopic description of the superconductor, only its proximity effect inside the covered nanowire core. This is done through the simplistic model $\Delta(\bm{r})=\Delta_0 \Theta(x)$, where $\Theta(x)$ is the Heaviside function.

%

To solve the above Hamiltonian, we calculate the electrostatic potential $\phi(\bm{r})$ taking into account all the different materials surrounding the wire and their dimensions. The computation implies solving self-consistently the Schr\"odinger eigenproblem with the Poisson equation, which reads
\begin{equation}
\bm{\nabla}\cdot\left(\epsilon(\bm{r})\bm{\nabla}\phi(\bm{r})\right)=-\rho_{\rm surf}(\bm{r})-\rho_{\rm TF}(\bm{r}).
\label{Eq:phi_Poisson}
\end{equation}
Here $\epsilon(\bm{r})$ is the inhomogeneous dielectric permittivity, which takes a different constant value inside each material. The charge density $\rho_{\rm surf}(\bm{r})$ models the charge accumulation layer that is typically present at the bare facets of this kind of semiconducting nanowires. We model it as a positive charge layer of $1$ nm thickness at the uncovered nanowire facets. $\rho_{\rm TF}(\bm{r})$ corresponds to the mobile charge inside the wire, that we compute in the Thomas-Fermi approximation
\begin{equation}
\rho_{\rm TF}(\bm{r})=-\frac{e}{3\pi^2}\left[\frac{2m^*\left|e\phi(\bm{r})\right|f(-e\phi(\bm{r}))}{\hbar^2}\right]^{\frac{3}{2}},
\label{Eq:rho_TF}
\end{equation}
where $f(\omega)=[1+\exp(\omega/k_BT)]^{-1}$ is the Fermi-Dirac distribution for a given energy $\omega$ and temperature $T$. We solve Eq. \eqref{Eq:phi_Poisson} together with Eq. \eqref{Eq:rho_TF} in a self-consistent manner
(see Ref.~\onlinecite{Escribano:PRB19, Escribano:20} for a detailed explanation of the self-consistent method in this precise context). We impose as boundary conditions a voltage $V_{\rm bg}$ at the back gate, a voltage $V_{\rm Al}$ at the boundary of the superconductor to simulate the band-bending between the wire and the superconductor, and a voltage $V_{\rm{N}}$ at the normal contact. We include a HfO$_2$ substrate of thickness $W_{\rm HfO}$ between the device and the back gate that is typically present in the experiments. Finally, we obtain the SOC in Eq. \eqref{Eq:H_3D} from the electrostatic potential for a (111) zinc-blende InAs nanowire, following the approach of Ref.~\onlinecite{Escribano:20},
\begin{eqnarray}
\bm{\alpha}(\bm{r})\simeq\frac{eP^2_{\rm fit}}{3}\left[\frac{1}{\left(\Delta_{\rm g}+e\phi(\bm{r})\right)^2} \right. \nonumber \\
\left. -\frac{1}{\left(\Delta_{\rm g}+\Delta_{\rm soff}+e\phi(\bm{r})\right)^2}\right]\cdot\bm{\nabla}\phi(\bm{r}).
\end{eqnarray}
Here, $\Delta_{\rm g}$ and $\Delta_{\rm soff}$ are the semiconductor and split-off band gaps of the semiconducting wire, and $P_{\rm fit}$ is the Kane parameter provided in Ref.~\onlinecite{Escribano:20} that includes in a phenomenological way the intravalence band couplings. All the parameters used in our simulations can be found in Table \ref{Table:Params_3D}.

\begin{table}[htb]
\centering
\caption{Parameters used for the microscopic model.}

\renewcommand{\arraystretch}{1.5}
\newcolumntype{m}{>{\hsize=0.5\hsize}X}
\newcolumntype{s}{>{\hsize=0.82\hsize}X}
\newcolumntype{l}{>{\hsize=1\hsize}X}
\newcolumntype{h}{>{\hsize=0.75\hsize}X}
\newcolumntype{j}{>{\hsize=1\hsize}X}

\begin{tabularx}{1\columnwidth}{ |>{\hsize=0.9\hsize}X|>{\hsize=0.9\hsize}X|>{\hsize=1.2\hsize}X| }
  \multicolumn{3}{c}{Hamiltonian} \\
  \hline \hline
  $m^*=0.023m_0$ & $\Delta_0=0.2$ meV & $P_{\rm fit}=1252$ meV$\cdot$nm \\ \hline
   $\Delta_{\rm g}=417$ meV & $\Delta_{\rm soff}=390$ meV & \\
  \hline
\end{tabularx}

\vspace*{0.22 cm}

\begin{tabularx}{1\columnwidth}{ |>{\hsize=0.9\hsize}X|>{\hsize=0.9\hsize}X|>{\hsize=1.2\hsize}X| }
  \multicolumn{3}{c}{Electrostatics} \\
  \hline \hline
  $\epsilon_{\rm InAs}=15.5\epsilon_0$ & $\epsilon_{\rm HfO}=25\epsilon_0$ & $\rho_{\rm acc}=2\cdot 10^{-3}\left(\frac{e}{\rm nm^3}\right)$ \\ \hline 
   $V_{\rm SC}=0.2$ eV & $V_{\rm N}=0$ & $T=10$ mK \\
  \hline
\end{tabularx}

\vspace*{0.22 cm}

\begin{tabularx}{1\columnwidth}{ |X|X|X| }
  \multicolumn{3}{c}{Geometrical} \\
  \hline \hline
  $R=60$ nm & $d=10$ nm & $L=800$ nm \\ \hline
  $L_{\rm S}=600$ nm & $W_{\rm HfO}=20$ nm & \\
  \hline
\end{tabularx}

\label{Table:Params_3D}
\end{table}

Once $\phi(\bm{r})$ and $\bm{\alpha}(\bm{r})$ are obtained for a particular back-gate potential $V_{\rm bg}$, we discretize the space inside the semiconducting nanowire into a rectangular grid with a $2$ nm lattice spacing, and then we transform the continuum Hamiltonian of Eq. \eqref{Eq:H_3D} into a tight-binding one. We diagonalize it to obtain the eigenenergies and eigenstates using the routines implemented in Ref.~\onlinecite{MajoranaNanowiresQSP_v1}.

In Fig. \ref{fig:SM0}(a) we show the energy spectrum $E$ versus the back-gate potential $V_{\rm bg}$. For $V_{\rm bg}\lesssim-0.3$ V  the energy spectrum exhibits a hard gap of $2\Delta_0=0.4$ meV, while for $V_{\rm bg}>-0.3$ V several subgap states appear. The spatial probability density of the typical subgap states for two $V_{\rm bg}$, marked with circles in (a), are plotted in Figs. \ref{fig:SM0}(d,e). The associated electrostatic potential is shown in Fig. \ref{fig:SM0}(f).
The weight $W_{\rm S}$ of the lowest-energy wavefunctions inside the covered region as a function of  $V_{\rm bg}$ is shown in Fig. \ref{fig:SM0}(b). Similarly, the expectation value $\langle z\rangle$ and its quadratic dispersion $\sigma_z^2$ in units of $R$ of the lowest state inside the covered region is shown is shown in Fig. \ref{fig:SM0}(c). These quantities allow us to give quantitative estimations to the parameters of the generalized single impurity Anderson (SIA) model, presented in the main text and discussed in detail in Appendix \ref{AppB}.


The first aspect to analyze is the spatial location of states. The potential $-e\phi(\bm r)$ is negative inside the full-shell region, particularly at the ohmic S interfaces (due to band bending), while it is closer to zero (Fermi energy) in the uncovered region. Hence, a QD state with energy below the gap will be located mostly in the uncovered region, with a part of its wavefunction of weight $W_{\rm S}$ inside the covered region ($x>0$). We see in Fig. \ref{fig:SM0}(b) that when the state energy is small in Fig. \ref{fig:SM0}(a), the associated weight $W_{\rm S}$ decreases but remains sizable (around 25\%). This is confirmed by the spatial wavefunction profiles [Figs. \ref{fig:SM0}(d,e)], which exhibit a large weight in the uncovered region, with a tail leaking along the S contact. This tail allows for a significant coupling of the state to the superconductor. 

The axial symmetry of the coupling is a second aspect that is crucial in the presence of fluxoid vortices, as discussed in the main text. It is microscopically determined by the symmetry of the state's wavefunction itself inside the covered $x>0$ region (the QD-S coupling can be expressed as an overlap of the QD and S wave functions at the interface between both materials). The symmetry of the wavefunction is estimated by the average position $\bm r_0 = \langle \bm r\rangle$ of its $x>0$ tail, so that $\bm r_0=0$ corresponds to a QD symmetrically coupled to the superconductor in the generalized SIA model. Since our device has $y\to-y$ symmetry, we have $\langle y\rangle =0$, and $|\bm r_0|\equiv r_0 = |\langle z\rangle|$.  Fig. \ref{fig:SM0}(c) shows that $\langle z\rangle$ remains very close to zero regardless of back gate voltage, particularly $r_0\lesssim 0.1R$. The ultimate reason for this strong axial-symmetry of the QD's tail inside the covered region is that the metallic shell completely screens the electrostatic potential created by the back-gate inside the covered region of the wire, as illustrated in Fig. \ref{fig:SM0}(f). Hence, the QD's tail retains the symmetry of the full-shell region regardless of the fact that the back-gate potential is located asymmetrically (only covering the bottom of the wire), and the fact that it may create a strongly asymmetric potential profile in the uncovered region. 
As a result, the QD-S coupling naturally tends to be axially symmetric in realistic situations. Notice that these conclusions are essentially independent of (reasonable) parameter variation as those in Table \ref{Table:Params_3D}, and are simply a direct consequence of the electrostatic screening inside the superconducting shell.

\begin{figure*}
   \centering
   \includegraphics[width=0.95\textwidth]{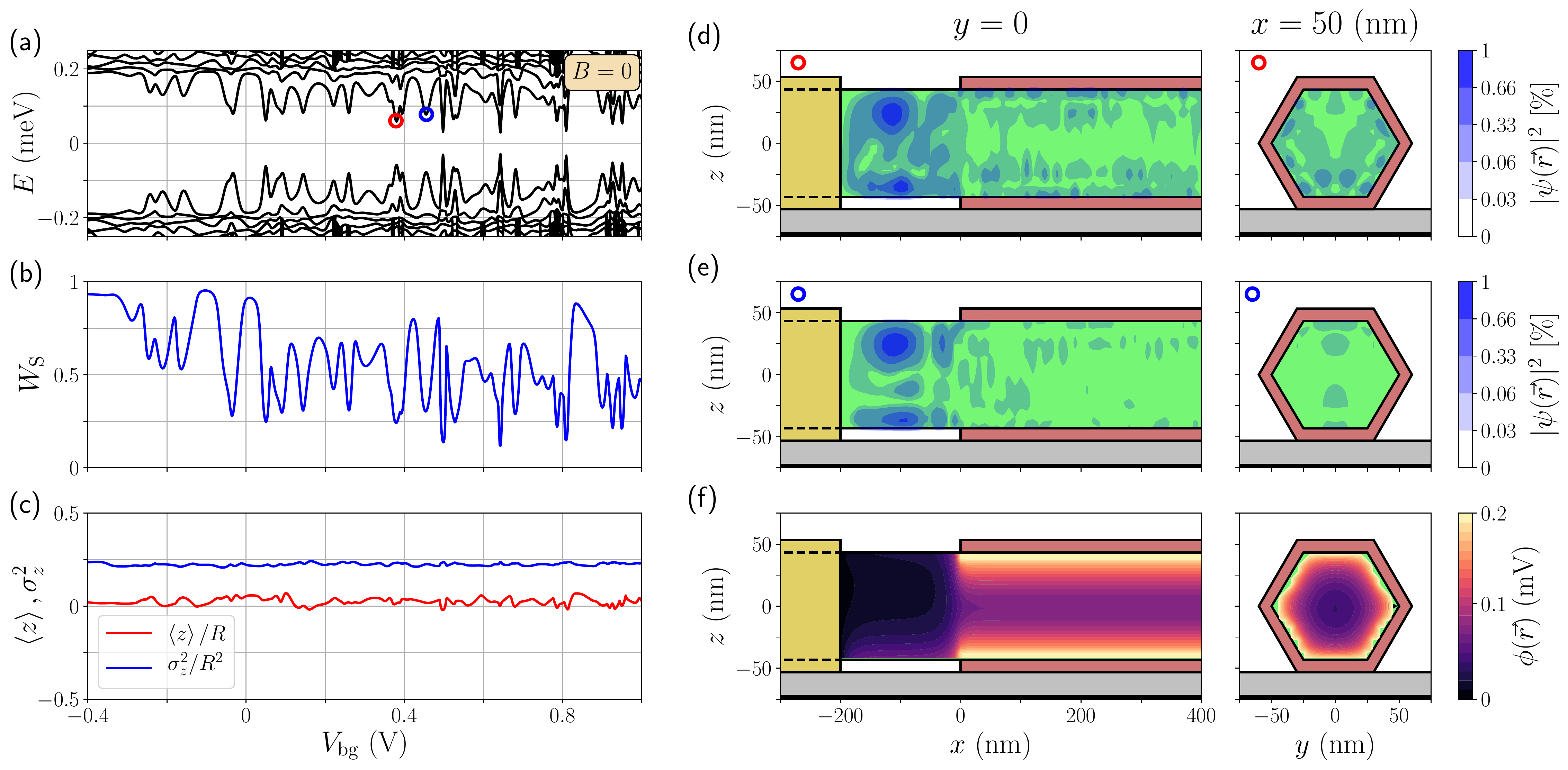}
   \caption{\textbf{Microscopic simulations.} (a) Energy spectrum at zero applied magnetic field of a finite-length device like the one in Fig. \ref{fig:1} of the main text, as a function of back-gate voltage $V_{\rm{bg}}$ (which is effective only in the uncovered segment of the wire). The level quantization is produced by lateral and longitudinal confinement. The presence of the electrostatic environment is taken into account at a self-consistent mean-field level. (b) Wave function weight $W_{\rm{S}}$ inside the shell-covered region of the wire vs $V_{\rm{bg}}$ for the lowest-energy eigenstate. (c) Expectation value of the $z$ position of the wavefunction inside the full-shell region $\langle z \rangle$ (red curve) and standard deviation $\sigma^2_z$ (blue curve) normalized to the wire's radius $R$ as a function of $V_{\rm{bg}}$ for the lowest-energy eigenstate.(d,e) Wavefunction profiles (probability density) inside the wire for the lowest energy level marked with a red/blue circle in (a). (f) Electrostatic profile inside the wire at $V_{\rm bg}=0.4$ V. Parameters are provided in Table \ref{Table:Params_3D}.}
   \label{fig:SM0}
\end{figure*}

\section{Generalized SIA model}
\label{AppB}

\subsection{Hamiltonian}
In the previous section, we described a 3D full-shell nanowire device microscopically (at $B=0$). In this section, we simplify the description, reducing it to its basic low-energy ingredients: a point like interacting QD state coupled on the right to a cylindrical S shell of finite radius $R$ and negligible thickness, and on the left to a one-dimensional normal lead. 

The presence of an axial external magnetic field, or alternatively a magnetic flux penetrating a thin-walled S cylinder, $\Phi=\pi R^2 B$, leads to three effects that need to be considered in the S-shell Hamiltonian: orbital effects, that are taking into account with a standard Peierls substitution, a winding of the superconducting pairing phase $\theta$ of the S order parameter $\bm\Delta = \Delta e^{-in\theta}$, where $n$ is the quantized fluxoid or fluxon \cite{Tinkham:04}, i.e., $n = \left \lfloor{\Phi/\Phi_0}\right \rceil$, with $\Phi_0=h/2e$ the S magnetic flux quantum, and a modulation of the S gap with the flux $\Delta(\Phi)$, explained in the next section. All these ingredients lead to a generalized superconducting impurity Anderson (SIA) model of the form
\begin{eqnarray}
H & = & H_{\rm D}+H_{\rm S}+H_{\rm SD}+H_{\rm N}+H_{\rm ND}, \\
H_{\rm D} & = & \sum_{\sigma}\left(\epsilon_{0}+\sigma V_{\rm Z}\right)d_{\sigma}^{\dagger}d_{\sigma} + Un_\uparrow n_\downarrow, \nonumber \\
H_{\rm S} & = & \sum_{\sigma,\theta,k_{S}}\left[\left(\epsilon_{k_{S}}-\mu_{\theta}\right)c_{\sigma,\theta,k_{S}}^{\dagger}c_{\sigma,\theta,k_{S}}\right. \nonumber \\
& & \left. -\tilde{t}_{\theta}\left(e^{i\frac{\pi}{N_\theta}\frac{\Phi}{\Phi_{0}}}c_{\sigma,\theta+1,k_{S}}^{\dagger}c_{\sigma,\theta,k_{S}} \right. \right. \nonumber \\
& & \left. \left. +e^{-i\frac{\pi}{N_\theta}\frac{\Phi}{\Phi_{0}}}c_{\sigma,\theta,k_{S}}^{\dagger}c_{\sigma,\theta+1,k_{S}}\right) \right. \nonumber \\
& & \left. + \Delta(\Phi)e^{-in\theta} c_{\sigma,\theta,k_{S}}^{\dagger}c_{-\sigma,\theta,-k_{S}}^{\dagger}+\rm{H.c.}\right], \nonumber \\
H_{\rm SD} & = & \sum_{\sigma,\theta,k_{S}}t(\theta)c_{\sigma,\theta,k_{S}}^{\dagger}d_{\sigma}+\rm{H.c.}, \nonumber \\
H_{\rm N} & = & \sum_{\sigma,k_{N}}\epsilon_{k_{N}}c_{\sigma,k_{N}}^{\dagger}c_{\sigma,k_{N}}, \nonumber \\
H_{\rm ND} & = & \sum_{\sigma,k_{N}}t_{N}c_{\sigma,k_{N}}^{\dagger}d_{\sigma}+\rm{H.c.} \nonumber
\label{Eq:Hamiltonan}
\end{eqnarray}
In this expression D, S and N denote the QD, the superconductor and the normal leads, correspondingly; and SD and ND the couplings between them. Spin-orbit coupling is neglected. The operator $d_{\sigma}$ corresponds to a single state in the QD, $c_{\sigma,\theta,k_S}$ to plane waves in the hollow cylinder localized at polar angle $\theta$ and with longitudinal momentum $k_S$, and $c_{\sigma,k_N}$ to states in the one-dimensional normal lead with longitudinal momentum $k_N$. The index $\sigma$ stands for the spin quantum number. The remaining parameters are introduced below with an explanation of each Hamiltonian. 

The spinful QD state is assumed to be point like. The parameter $\epsilon_0$ in $H_{\rm D}$ is the unperturbed energy of the state, while $V_{\rm Z}=\frac{1}{2}g\mu_{\rm B}B$ is its Zeeman splitting, being $g$ the g-factor of the nanowire and $\mu_B$ the Bohr magneton. This Zeeman field is due to the external magnetic field $B$ applied along the wire's direction $x$ [see sketch of Fig. \ref{fig:1}(a) of the main text]. In addition to this, we model the electron-electron interactions inside the QD with a charging energy $U$. We treat the interaction term at a mean-field level in the so-called Hartree-Fock-Bogoliubov approximation (more details on this approximation are explained in Appendix \ref{Subsec:numerical_sol}).

The normal lead is simply described as a semi-infinite one-dimensional chain with an energy dispersion relation $\epsilon_{k_N}$ (we neglect the Zeeman splitting in both the S and N leads), which is coupled at its end to the QD through hopping amplitude $t_{\rm N}$. In contrast, we describe the superconductor as a semi-infinite hollow cylinder of radius $R$. We discretize the cylinder into $N_\theta=\frac{2\pi}{\delta_\theta}$ one-dimensional semi-finite chains, where $\delta_\theta$ is the discretization spacing along the angular direction. Each chain ends at the cylinder rim, at position $\bm r_{\theta_i} = (0, R\cos\theta_i, R\sin\theta_i)$, where $\theta_i/\delta_\theta = 1,\dots,N_\theta$, and is only coupled to its two neighboring chains. We consider the interchain coupling $\tilde{t}_\theta=\frac{\hbar^2}{2m_{\rm S}R^2\delta_\theta^2}$ to be independent of the angle ($m_{\rm S}$ is the effective S mass). This hopping between chains shifts the (normal) energy dispersion relation of each chain $\epsilon_{k_S}$ by a quantity $\mu_{\theta}=-2\tilde{t}_\theta$ in the above Hamiltonian. Note that $\epsilon_{-k_{S}}=\epsilon_{k_{S}}$. 

\subsection{Ginzburg-Landau solution of the Little-Parks effect}

The applied magnetic flux $\Phi=\pi R^2 B$ leads to a modulation of the superconducting gap $\Delta(\Phi)$ according to the well-known Little-Parks (LP) effect~\cite{Little:PRL62, Tinkham:04}. The LP phenomenon can be described phenomenologically using the Ginzburg-Landau formalism for a superconductor in the presence of a magnetic field. Assuming a ballistic superconductor whose size is much smaller than the superconducting coherence length, the solution to the Ginzburg-Landau equations gives rise to the following equations \cite{Bardeen:RMP62, Larkin:05, Valentini:S21}
\begin{eqnarray}
\frac{\Delta(\Phi)}{\Delta_0}\approx \left(\frac{T_{\rm C}(\Phi)}{T_{\rm C}^0}\right)^{\frac{3}{2}}, \nonumber \\
\ln\left(\frac{T_{\rm C}(\Phi)}{T_{\rm C}^0}\right) +\mathcal{D}\left(\frac{1}{2}+\frac{\Lambda_n(\Phi)}{2\pi T_{\rm C}{\Phi}}\right)-\mathcal{D}\left(\frac{1}{2}\right)=0, 
\label{Eq:SC_gap}
\end{eqnarray}
where $\Delta(\Phi)$ is the S gap modulated by the magnetic flux [being $\Delta_0=\Delta(0)$ the S gap at zero magnetic field], $T_{\rm C}(\Phi)$ the flux-dependent critical temperature (being $T_{\rm C}^0$ at zero magnetic field), and $\mathcal{D}(x)$ and $\Lambda_n(\Phi)$ the standard digamma function and the pair-breaking term of a hollow superconducting cylinder, respectively. The latter is also derived in the Ginzburg-Landau formalism in the presence of impurities~\cite{Groff:PR68}
\begin{eqnarray}
\Lambda_n(\Phi)=\frac{T_{\rm C}^0}{\pi}\frac{\xi^2}{R^2}\left[4\left(n-\frac{\Phi}{\Phi_0}\right)^2 \right. \nonumber \\
\left. +\frac{d^2}{R^2}\left(\frac{\Phi^2}{\Phi_0^2} +\left(\frac{1}{3}+\frac{d^2}{20R^2}\right)n^2\right)\right].
\label{Eq:Lambda_n}
\end{eqnarray}
Here, $\Phi_0=h/2e$ is the S magnetic flux quantum, $\xi$ is the superconducting coherence length (at zero magnetic field) and $R$ and $d$ are the radius and thickness of the hollow cylinder, respectively. We note that Eq. \eqref{Eq:SC_gap} does not have an exact analytical solution in general, and thus it must be solved numerically. In the limit of $d\rightarrow 0$, though, an approximated solution exists \cite{Vaitiekenas:S20}
\begin{equation}
\Delta(\Phi)\approx \Delta_0\max\left[1-\frac{\xi^2}{R^2}\left(n-\frac{\Phi}{\Phi_0}\right)^2, 0\right].
\end{equation}

\subsection{Continuum limit for the cylinder}

The S Hamiltonian $H_{\rm S}$ can be written in the continuum limit $\delta_\theta\rightarrow 0$, which makes it quadratic in the orbital angular momentum operator $k_\theta = -i\partial_\theta$. Since the superconductor is a cylinder, we must impose periodic boundary conditions, and therefore the orbital angular momentum can only take integer values $m$. 
The S Hamiltonian becomes
\begin{eqnarray}
H_{\rm S} & = & \sum_{\sigma,m,k_{S}}\left[\left(\epsilon_{k_{S}}+t_\theta\left(m+\frac{\Phi}{2\Phi_0}\right)^2\right)c_{\sigma,m,k_{S}}^{\dagger}c_{\sigma,m,k_{S}} \right. \nonumber \\
& & \left. + \Delta(\Phi) c_{\sigma,m,k_{S}}^{\dagger}c_{-\sigma,-n-m,-k_{S}}^{\dagger}+\rm{H.c.}\right],
\end{eqnarray}
where we define $t_\theta=\tilde{t}_\theta\left(\frac{2\pi}{N_\theta}\right)^2=\frac{\hbar^2}{2m_{\rm S}R^2}$, which is also independent of the angle. Notice that the Hamiltonian is not block-diagonal in the $(m,-m)$ Nambu sectors, but in the $(m,-m-n)$ ones, i.e., one can write
\begin{widetext}
\begin{eqnarray}
H_{S} & = & \frac{1}{2}\sum_{\sigma,m,k_S}\left(\begin{array}{cc}
c_{\sigma,m,k_S}^{\dagger} & c_{-\sigma,-m-n,-k_S}\end{array}\right)\tilde{H}_{\sigma,m,k_S}^{S}\left(\begin{array}{c}
c_{\sigma,m,k_S}\\
c_{-\sigma,-m-n,-k_S}^{\dagger}
\end{array}\right), \\
\tilde{H}_{\sigma,m,k_S}^{S} & = & \left(\begin{array}{cc}
\epsilon_{k_{S}}+t_{\theta}\left(m+\frac{\Phi}{2\Phi_{0}}\right)^{2} & \Delta(\Phi)\\
\Delta(\Phi) & -\epsilon_{k_{S}}-t_{\theta}\left(m+n-\frac{\Phi}{2\Phi_{0}}\right)^{2}
\end{array}\right). \nonumber
\end{eqnarray}
\end{widetext}
This means that time-reversal partners $m$ and $-m$ are not coupled anymore, except for the zeroth LP lobe $n=0$, which is actually a special case in this work.

\subsection{Modeling the QD-S coupling}
\label{Sec:QD-S_coupling}
Finally, we describe the coupling between the QD and the superconductor through the following hopping parameter
\begin{equation}
t(\theta)=t_{\rm S}\exp\left(-\beta\frac{\left|\bm{r}_0-\bm{r}_\theta\right|}{R}\right),
\label{Eq:t(theta)}
\end{equation}
where $\bm{r}_0=\left(0, \left<y\right>,\left<z\right>\right)$ is the (point like QD position, microscopically representing the wavefunction mean position in the region inside the shell), $\bm{r}_\theta=R(0,\cos\theta,\sin\theta)$ is the end of each $\theta$-chain along the rim of the S cylinder, and $\beta$ is a dimensionless parameter that quantifies the degree of overlap between the QD and the S states. Since we are describing the complex QD wave function as a single site, in the spirit of the tight-binding model, we may assume that $\beta$ is roughly inversely proportional to the QD wave function width given by the wavefunction standard deviation.


\subsection{QD self energies}

We are interested in the spectral observables of the QD. These can be obtained from the QD retarded Green's function for each spin $\sigma$, given by
\begin{equation}
G^{\rm D}_\sigma(\omega,\Phi)=\left[G^{\rm D, 0}_\sigma(\omega,\Phi)^{-1}-\Sigma_\sigma(\omega,\Phi)\right]^{-1},
\end{equation}
where 
\begin{eqnarray}
\label{G0}
G^{\rm D, 0}_\sigma(\omega,\Phi)^{-1} = \left(\begin{array}{cc}
\omega-\epsilon_0-\sigma V_{\rm Z} & 0 \\
0  & \omega+\epsilon_0-\sigma V_{\rm Z} 
\end{array}\right)
\end{eqnarray}
is the inverse of the bare QD Green's function for spin $\sigma$ (note the Zeeman field $V_Z$ on the dot), and 
\begin{equation}
\Sigma_\sigma(\omega,\Phi)=\Sigma^{\rm U}_\sigma(\omega,\Phi)+\Sigma^{\rm S}(\omega,\Phi)+\Sigma^{\rm N}
\end{equation}
is the total self-energy. It includes the effect of Coulomb interactions in the QD, $\Sigma^{\rm U}_\sigma(\omega,\Phi)$, the coupling to the superconductor $\Sigma^{\rm S}(\omega,\Phi)$, and the coupling to the normal lead, $\Sigma^{\rm N}$. The later is merely given by the tunneling rate $\Sigma_{\rm N}=-i\Gamma_N=-i\pi|t_{\rm N}|^2\rho_{\rm N}$, being $\rho_{\rm N}$ the density of states of the normal lead at the Fermi energy. On the other hand, we treat the Coulomb interactions in the QD in the Hartree-Fock-Bogoliubov approximation, which constitutes the lowest $U$-order diagrammatic expansion of the QD self-energy
\begin{equation}
\Sigma^U_\sigma(\omega,\Phi)\approx U \left(\begin{array}{cc}
\left<n_{-\sigma}\right> & \left<d_\sigma d_{-\sigma}\right> \\
\left<d^\dagger_\sigma d^\dagger_{-\sigma}\right> & -\left<n_{\sigma}\right> 
\end{array}\right).
\label{Eq:Self_energy_HFB}
\end{equation}
These expectation values can be computed from the QD Green's function in the following way
\begin{eqnarray}
\left\langle n_{\sigma}\right\rangle =-\frac{1}{2\pi}\int d\omega \ \mathrm{Im}\left\{ \left(\mathrm{G^{\rm D}_\sigma\left(\omega,\Phi\right)}\right)_{00}f(\omega) \right. \nonumber \\
\left. + \left(\mathrm{G^{\rm D}_{-\sigma}\left(\omega,\Phi\right)}\right)_{11}f(-\omega)\right\} , \label{Eq:occupation} \\
\left\langle d_{\sigma}d_{\sigma}\right\rangle =\left\langle d_{\sigma}^{\dagger}d_{\sigma}^{\dagger}\right\rangle = \nonumber\\ 
-\frac{1}{\pi}\int d\omega \ \mathrm{Im}\left\{  \left(\mathrm{G^{\rm D}_\sigma\left(\omega,\Phi\right)}\right)_{01}f(\omega)\right\},
\end{eqnarray}
where $f(\omega)=[1+\exp(\omega/k_BT)]^{-1}$ is the Fermi-Dirac distribution for a given energy $\omega$ and temperature $T$. Note that these equations require a self-consistent solution. We discuss the self-consistency process further in Sec. \ref{Subsec:numerical_sol}. Finally, the self-energy due to the coupling with the superconductor is obtained by performing the following integrals
\begin{equation}
\Sigma^{\rm S}(\omega,\Phi)=\int dk_{\rm S} \int d\theta d\theta' t(\theta) g^{\rm S}(\omega,\Phi;\theta,\theta')t(\theta'),
\label{Eq:Sigma_S1}
\end{equation}
where 
\begin{equation}
g^{\rm S}(\omega,\Phi;\theta,\theta')=\mathcal{F}\left\{g^{\rm S}(\omega,\Phi;m)\right\}
\end{equation}
is the Fourier transform of the retarded Green's function of the decoupled S shell around its rim, which can in turn be written as
\begin{widetext}
\begin{eqnarray}
g^{\rm S}(\omega,\Phi;m) & = & \frac{1}{D_{m,n}}\left(\begin{array}{cc}
\omega+\left(\epsilon_{k_{S}}-L_{m}(\Phi)\right)-\Phi_{n}+2\sqrt{\Phi_{n}L_{m}} & \Delta(\Phi)\\
\Delta(\Phi) & \omega-\left(\epsilon_{k_{S}}-L_{m}(\Phi)\right)
\end{array}\right),
\end{eqnarray}
\end{widetext}
where we define
\begin{eqnarray}
L_{m}(\Phi)\equiv-t_{\theta}\left(m+\frac{\Phi}{2\Phi_{0}}\right)^{2}, \\
\Phi_{n}(\Phi)\equiv-t_{\theta}\left(n-\frac{\Phi}{\Phi_{0}}\right)^{2},
\label{Eq:Phi_and_L}
\end{eqnarray}
and
\begin{eqnarray}
D_{m,n}\equiv\omega^{2}-\left(\epsilon_{k_{S}}-L_{m}\right)^{2}-\Delta(\Phi)^{2} \nonumber \\
-\left[\omega-\left(\epsilon_{k_{S}}-L_{m}\right)\right]\left(\Phi_{n}-2\sqrt{\Phi_{n}L_{m}}\right)
\end{eqnarray}
is the determinant of the inverse of the Green's function. Its Fourier transform is
\begin{widetext}
\begin{equation}
g^{\rm S}(\omega,\Phi;\theta,\theta')=\sum_{m}\frac{e^{-im(\theta-\theta')}}{D_{m,n}}\left(\begin{array}{cc}
\left[\omega+\left(\epsilon_{k_{S}}-L_{m}(\Phi)\right)-\Phi_{n}+2\sqrt{\Phi_{n}L_{m}}\right] & \Delta(\Phi) e^{in\theta'}\\
\Delta(\Phi) e^{-in\theta} & \left[\omega-\left(\epsilon_{k_{S}}-L_{m}(\Phi)\right)\right]e^{-in(\theta-\theta')}
\end{array}\right).
\end{equation}
Now we perform the integrals in Eq. \eqref{Eq:Sigma_S1} to obtain the self-energy. To simplify them, we can expand the coupling $t(\theta)$ of Eq. \eqref{Eq:t(theta)} in a Fourier series. Without loss of generality we assume that the point like dot position is $\bm{r}_0=(0, r_0,0)$, which results in
\begin{equation}
t(\theta)=\sum_{m=0}^\infty t_m \cos(m\theta),
\end{equation}
where $m$ is a positive integer. The Fourier coefficients 
\begin{eqnarray}
t_0=\frac{1}{2\pi}\int_{-\pi}^{\pi}t(\theta)d\theta, \\  t_m=\frac{1}{\pi}\int_{-\pi}^{\pi}t(\theta)\cos(m\theta)d\theta\leftarrow\forall m>0,
\end{eqnarray}
can be computed numerically for any $R$ and $\beta$. Performing the integrals on $\theta$ and $\theta'$, we obtain the following Nambu elements for the self-energy:
\begin{eqnarray}
\Sigma^{\rm S}_{00}(\omega,\Phi) & = & \int dk_{S}\left(2\pi t_{0}\right)^{2}\frac{\left[\omega+\left(\epsilon_{k_{S}}-L_{0}(\Phi)\right)-\Phi_{n}+2\sqrt{\Phi_{n}L_{0}}\right]}{D_{0,n}}\nonumber \\
 & + & \sum_{m=1}^{\infty}\int dk_{S}\left(\pi t_{m}\right)^{2}\left\{ \frac{\left[\omega+\left(\epsilon_{k_{S}}-L_{m}(\Phi)\right)-\Phi_{n}+2\sqrt{\Phi_{n}L_{m}}\right]}{D_{m,n}} \right. \nonumber \\
  &  & \left. +\frac{\left[\omega+\left(\epsilon_{k_{S}}-L_{-m}(\Phi)\right)-\Phi_{n}+2\sqrt{\Phi_{n}L_{-m}}\right]}{D_{-m,n}}\right\} ,\\
\Sigma^{\rm S}_{11}(\omega,\Phi) & = & \int dk_{S}\left(2\pi t_{0}\right)^{2}\frac{\left[\omega-\left(\epsilon_{k_{S}}-L_{-n}(\Phi)\right)\right]}{D_{-n,n}}\nonumber \\
 & + & \sum_{m=1}^{\infty}\int dk_{S}\left(\pi t_{m}\right)^{2}\left\{ \frac{\left[\omega-\left(\epsilon_{k_{S}}-L_{m-n}(\Phi)\right)\right]}{D_{m-n,n}}+\frac{\left[\omega-\left(\epsilon_{k_{S}}-L_{-m-n}(\Phi)\right)\right]}{D_{-m-n,n}}\right\} , \\
\Sigma^{\rm S}_{10}(\omega,\Phi) & = & \Sigma^{\rm S}_{01}(\omega,\Phi)^{*},\\
\Sigma^{\rm S}_{01}(\omega,\Phi) & = & \delta_{n,0}\left\{ \int dk_{S}\left(2\pi t_{0}\right)^{2}\frac{\Delta(\Phi)}{D_{0,n}}+\sum_{m=1}^{\infty}\left(\pi t_{m}\right)^{2}\left[\frac{\Delta(\Phi)}{D_{m,n}}+\frac{\Delta(\Phi)}{D_{-m,n}}\right]\right\} \nonumber \\
 & + & \left(1-\delta_{n,0}\right)\left\{ \int dk_{S}\left(2\pi t_{0}\right)^{2}\frac{t_{n}}{2t_{0}}\frac{\Delta(\Phi)}{D_{0,n}}\right.\nonumber \\
 &  & \left. +\sum_{m=1}^{\infty}\left(1-\delta_{n,m}\right)\left(2\pi t_{0}\right)^{2}\frac{t_{m}t_{m-n}}{2t_{0}^{2}}\frac{\Delta(\Phi)}{D_{m-n,n}}+\left(1-\delta_{n,-m}\right)\left(2\pi t_{0}\right)^{2}\frac{t_{m}t_{m+n}}{2t_{0}^{2}}\frac{\Delta(\Phi)}{D_{-m-n,n}}\right\}.
\end{eqnarray}
Finally, to perform the integral on $k_{\rm S}$, we assume that the Fermi energy is the largest energy scale of the system (wide band limit), as is the case for Al, whose Fermi energy is $\sim 10$ eV~\cite{AschroftMermin:76}. This allows us to write $\epsilon_{k_S}\approx\hbar v_F k_{S}$ around the Fermi energy and transform the integral $\int dk_S\rightarrow \rho_{\rm S}\int d\epsilon_{k_{\rm S}}$, where $\rho_{\rm S}$ is the density of states of the S hollow cylinder at the Fermi level. The integrals under this approximation provide the following expressions [substituting $\Phi_n$ and $L_m$ by their expressions in Eq. \eqref{Eq:Phi_and_L}]:
\begin{eqnarray}
\Sigma^{\rm S}_{00}(\omega,\Phi) =-\Gamma_{\rm S}^{(0)}\frac{\omega+\frac{n}{2}t_{\theta}\left(n-\frac{\Phi}{\Phi_{0}}\right)}{\sqrt{\Delta(\Phi)^{2}-\left(\omega+\frac{n}{2}t_{\theta}\left(n-\frac{\Phi}{\Phi_{0}}\right)\right)^{2}}}\nonumber \\
-\sum_{m=1}^{\infty}\Gamma_{\rm S}^{(m)}\left\{ \frac{\omega+\left(\frac{n}{2}+m\right)t_{\theta}\left(n-\frac{\Phi}{\Phi_{0}}\right)}{\sqrt{\Delta(\Phi)^{2}-\left(\omega+\left(\frac{n}{2}+m\right)t_{\theta}\left(n-\frac{\Phi}{\Phi_{0}}\right)\right)^{2}}}+\frac{\omega+\left(\frac{n}{2}-m\right)t_{\theta}\left(n-\frac{\Phi}{\Phi_{0}}\right)}{\sqrt{\Delta(\Phi)^{2}-\left(\omega+\left(\frac{n}{2}-m\right)t_{\theta}\left(n-\frac{\Phi}{\Phi_{0}}\right)\right)^{2}}}\right\} ,\label{Eq:Self_energy_SC1} \\
\Sigma^{\rm S}_{11}(\omega,\Phi)=-\Gamma_{\rm S}^{(0)}\frac{\omega-\frac{n}{2}t_{\theta}\left(n-\frac{\Phi}{\Phi_{0}}\right)}{\sqrt{\Delta(\Phi)^{2}-\left(\omega-\frac{n}{2}t_{\theta}\left(n-\frac{\Phi}{\Phi_{0}}\right)\right)^{2}}}\nonumber \\
-\sum_{m=1}^{\infty}\Gamma_{\rm S}^{(m)}\left\{ \frac{\omega-\left(\frac{n}{2}-m\right)t_{\theta}\left(n-\frac{\Phi}{\Phi_{0}}\right)}{\sqrt{\Delta(\Phi)^{2}-\left(\omega-\left(\frac{n}{2}-m\right)t_{\theta}\left(n-\frac{\Phi}{\Phi_{0}}\right)\right)^{2}}}+\frac{\omega-\left(\frac{n}{2}+m\right)t_{\theta}\left(n-\frac{\Phi}{\Phi_{0}}\right)}{\sqrt{\Delta(\Phi)^{2}-\left(\omega-\left(\frac{n}{2}+m\right)t_{\theta}\left(n-\frac{\Phi}{\Phi_{0}}\right)\right)^{2}}}\right\} ,
\end{eqnarray}
\begin{eqnarray}
\Sigma^{\rm S}_{10}(\omega,\Phi)=\Sigma^{\rm S}_{01}(\omega,\Phi)^{*},\\
\Sigma^{\rm S}_{01}(\omega,\Phi)=-\delta_{n,0}\left\{ \Gamma_{\rm S}^{(0)}\frac{\Delta(\Phi)}{\sqrt{\Delta(\Phi)^{2}-\left(\omega+\frac{n}{2}t_{\theta}\left(n-\frac{\Phi}{\Phi_{0}}\right)\right)^{2}}}\right.\nonumber \\
\left.+\sum_{m=1}^{\infty}\Gamma_{\rm S}^{(m)}\left[\frac{\Delta(\Phi)}{\sqrt{\Delta(\Phi)^{2}-\left(\omega+\left(\frac{n}{2}+m\right)t_{\theta}\left(n-\frac{\Phi}{\Phi_{0}}\right)\right)^{2}}}+\frac{\Delta(\Phi)}{\sqrt{\Delta(\Phi)^{2}-\left(\omega+\left(\frac{n}{2}-m\right)t_{\theta}\left(n-\frac{\Phi}{\Phi_{0}}\right)\right)^{2}}}\right]\right\} \nonumber \\
-\left(1-\delta_{n,0}\right)\Gamma_{\rm S}^{(0)}\left\{ \sqrt{\frac{\Gamma_{\rm S}^{(n)}}{\Gamma_{\rm S}^{(0)}}}\frac{\Delta(\Phi)}{\sqrt{\Delta(\Phi)^{2}-\left(\omega+\frac{n}{2}t_{\theta}\left(n-\frac{\Phi}{\Phi_{0}}\right)\right)^{2}}}\right.\nonumber \\
\left.+\sum_{m=1}^{\infty}\left(1-\delta_{m,n}\right)2\frac{\sqrt{\Gamma_{\rm S}^{(m)}\Gamma_{\rm S}^{(m-n)}}}{\Gamma_{\rm S}^{(0)}}\frac{\Delta(\Phi)}{\sqrt{\Delta(\Phi)^{2}-\left(\omega-\left(\frac{n}{2}-m\right)t_{\theta}\left(n-\frac{\Phi}{\Phi_{0}}\right)\right)^{2}}}\right.\nonumber \\
\left.+\left(1-\delta_{m,-n}\right)2\frac{\sqrt{\Gamma_{\rm S}^{(m)}\Gamma_{\rm S}^{(m+n)}}}{\Gamma_{\rm S}^{(0)}}\frac{\Delta(\Phi)}{\sqrt{\Delta(\Phi)^{2}-\left(\omega-\left(\frac{n}{2}+m\right)t_{\theta}\left(n-\frac{\Phi}{\Phi_{0}}\right)\right)^{2}}}\right\}, \label{Eq:Self_energy_SC2}
\end{eqnarray}
\end{widetext}
where we have introduced the tunneling rate components $\Gamma_{\rm S}^{(m)}$ as
\begin{eqnarray}
\Gamma_{\rm S}^{(0)}\equiv \pi\rho_{\rm S}(2\pi t_0)^2=\pi\rho_{\rm S}\left(\int_{-\pi}^{\pi}t(\theta)d\theta\right)^2, \\
\Gamma_{\rm S}^{(m)}\equiv \pi\rho_{\rm S}(\pi t_m)^2 \nonumber \\ =\pi\rho_{\rm S}\left(\int_{-\pi}^{\pi}t(\theta)\cos(m\theta)d\theta\right)^2.
\end{eqnarray}
For the particular case of Eq. \eqref{Eq:t(theta)} we can actually write
\begin{eqnarray}
\Gamma_{\rm S}^{(0)}=\Gamma_{\rm S}\left(\int_{-\pi}^{\pi}\exp\left(-\beta\frac{\left|\bm{r}_0-\bm{r}_\theta\right|}{R}\right)d\theta\right)^2, \\
 \Gamma_{\rm S}^{(m)} \nonumber \\ 
 =\Gamma_{\rm S}\left(\int_{-\pi}^{\pi}\exp\left(-\beta\frac{\left|\bm{r}_0-\bm{r}_\theta\right|}{R}\right)\cos(m\theta)d\theta\right)^2,
\end{eqnarray}
where $\Gamma_{\rm S}=\pi|t_{\rm S}|^2\rho_{\rm S}$ is the mean tunneling rate to the superconductor and one of the parameters that we tune in our simulations.

Equations  \eqref{Eq:Self_energy_SC1}-\eqref{Eq:Self_energy_SC2} are the final form of the self-energy due to the superconductor that we use in our simulations. They show that the QD is in general coupled to all the different angular momentum numbers $m$ of the superconductor through the tunnel rate components $\Gamma_{\rm S}^{(m)}$ (or alternatively $t_m$). For the coupling $t(\theta)$ of Eq. \eqref{Eq:t(theta)} proposed in this work, these components decay fast with $m$, what allows us to impose a cut-off in the summations. Particularly, in our simulations we only sum the first $20$ components. Another aspect that is worth mentioning is that $\Sigma^{\rm S}_{00}(\omega,\Phi;m)\neq -\Sigma^{\rm S}_{11}(\omega,\Phi;m)$, except for $n=0$ or $\omega\rightarrow 0$. The reason, as mentioned previously, is the fact that the time-reversal partners for $n\ge 1$ lobes are $m$ and $-n-m$. Therefore the fluxon number $n$ becomes mixed with the angular momentum of the superconductor, giving rise to a generalized angular momentum. Additionally, the winding of the phase in the superconductor gives rise to a different anomalous self-energy depending on whether we drive the system to the zeroth ($n=0$) or subsequent ($n\neq 0$) LP lobes. In the later case, these equations also shows that if the QD is placed on the cylinder axis, and therefore $\Gamma_{\rm S}^{(m)}= 0\leftarrow\forall m>0$, then the pairing induced on the QD is exactly zero. In this particular case, if we also assume that $t_{\theta}\ll \omega,\Delta$ or that $\Phi = n\Phi_0$, we can write
\begin{eqnarray}
\Sigma^{\rm S}(\omega,\Phi)\approx -\frac{\Gamma_{\rm S}^{(0)}}{\sqrt{\Delta^2-\omega^2}}\left(\begin{array}{cc}
\omega & \Delta\delta_{n,0} \\
\Delta\delta_{n,0}  & \omega 
\end{array}\right),
\end{eqnarray}
which is the canonical self-energy of a singly connected S lead in the wide band limit, except for the Kronecker delta $\delta_{n,0}$ in the off-diagonal elements. As mentioned in the main text, a similar  $\Sigma^{\rm S}$ arises in S-QD-S Josephson junctions at phase difference $\pi$, where the pairing also vanishes when summing the contribution of two symmetrically coupled S leads.

\subsection{Numerical solutions}
\label{Subsec:numerical_sol}
We want to analyze the local density of states (LDOS) of our system at the QD, $\rho(\omega,\Phi)$. It can be computed from the retarded Green's function of the QD
\begin{equation}
\rho(\omega,\Phi)=-\frac{1}{\pi}\sum_\sigma \mathrm{Im}\left\{\mathrm{Tr}\left\{G^{\rm D}_\sigma(\omega,\Phi)\right\}\right\}.
\end{equation}
To this end, we first solve self-consistently Eqs. \eqref{Eq:SC_gap} and \eqref{Eq:Lambda_n} to obtain the superconducting gap $\Delta(\Phi)$ modulated by the LP effect, and then we write numerically the QD Green's function, which as shown before is
\begin{equation}
G^{\rm D}_\sigma(\omega,\Phi)=\left[G^{\rm D, 0}_\sigma(\omega,\Phi)^{-1} -\Sigma_\sigma(\omega,\Phi)\right]^{-1}.
\label{Eq:G_D}
\end{equation}
The inverse of the bare QD Green's function is given by Eq. \eqref{G0}. The self-energy due to the N lead is $\Sigma_{\rm N}(\sigma)=-i\Gamma_N$, and the self-energy terms that takes into account the Coulomb interaction and the coupling to the superconductor are given by Eq. \eqref{Eq:Self_energy_HFB} and Eqs. (\ref{Eq:Self_energy_SC1}-\ref{Eq:Self_energy_SC2}), respectively. Note that one must obtain the QD Green's function of Eq. \eqref{Eq:G_D} in a self-consistent manner, since it depends on the occupation $\left<n_\sigma\right>$, which in turn depends on $G^{\rm D}_\sigma(\omega,\Phi)$. To solve this self-consistency, we use an Anderson mixing iterative solver with self-adaptive coefficients. Within this approach, we first compute $G^{\rm D}_\sigma(\omega,\Phi)$ of Eq. \eqref{Eq:G_D} assuming a seed value for the occupations $\left<n_\sigma\right>$ and $\left<d_\sigma d_{-\sigma}\right>$. We take as a starting point $\left<n_\uparrow\right>=1$ and $\left<n_\downarrow\right>=0$, being "$\uparrow$" the direction in which the magnetic field is applied, and $\left<d_\sigma d_{-\sigma}\right>=0$. Then we obtain all the expectation values $\left< n_\sigma\right>$ and $\left<d_\sigma d_{-\sigma}\right>$ from $G^{\rm D}_\sigma(\omega,\Phi)$. Next we compute the self-energy due to the Coulomb interaction, but instead of simply using the newly computed occupations $\left<n_\sigma\right>_{(i)}$ and $\left<d_\sigma d_{-\sigma}\right>_{(i)}$, we mix them with the ones obtained in the preceding $i-1$ step. In other words, we perform the replacement
\begin{eqnarray}
\left<n_\sigma\right>_{(i)} &\to& (1-\gamma)\left<n_\sigma\right>_{(i)}+\gamma\left<n_\sigma\right>_{(i-1)}, \\
\left< d_\sigma d_{-\sigma}\right>_{(i)} &\to& (1-\gamma)\left<d_\sigma d_{-\sigma}\right>_{(i)}+\gamma\left<d_\sigma d_{-\sigma}\right>_{(i-1)}, \nonumber
\label{Eq:Anderson_mixing}
\end{eqnarray}
where $\gamma\in \left[0,1\right]$ is the so-called Anderson mixing coefficient. We repeat this procedure until sufficient convergence between two consecutive steps is reached. Particularly, in our simulations, we stop the procedure when $|\left<n_\sigma\right>_{(i)}-\left<n_\sigma\right>_{(i-1)}|\le 10^{-6}$ for both spin directions. The mixing shown above ensures a stable convergence of the system if $\gamma$ is small enough. However, a too small $\gamma$ slows the convergence of the system. To overcome this problem, we use a self-adaptive coefficient that we allow to vary depending on the convergence speed:
\begin{equation}
\gamma=\gamma_{\rm max} \cdot \exp\left(-2 \frac{|\left<n_\sigma\right>_{(i)}-\left<n_\sigma\right>_{(i-1)}|}{\max\left\{|\left<n_\sigma\right>_{(i)}|,|\left<n_\sigma\right>_{(i-1)}|\right\}} \right),
\end{equation}
where here $\left<n_\sigma\right>_{(i)}$ is the occupation \emph{before} mixing, and $\gamma_{\rm max}$ is the maximum value that we allow $\gamma$ to take ($\gamma_{\rm max}=0.1$ in most of our simulations).

There are some other observables of interest that can be computed with $G^{\rm D}_\sigma(\omega,\Phi)$. The magnetization $P=\left<n_\uparrow\right>-\left<n_\downarrow\right>$, calculated from Eq. \eqref{Eq:occupation}, shows whether the QD is in a singlet ($P=0$) or doublet phase ($\left|P\right|>0$). In addition, the zeros of the determinant of the inverse of the QD Green's function, $\mathrm{det}\left\{G^{\rm D}_\sigma(\omega)^{-1}\right\}=0$, provide the eigenvalues of the QD Hamiltonian $E$. 

A summary with all the parameters we use for the generalized SIA model can be found in Table \ref{Table:Params_Anderson}. Notice that $\epsilon_0$, $B$, $\Gamma_{\rm S}$ and $r_0$ are not given because they are free parameters in our work (they can actually be tuned experimentally).

\begin{figure*}
   \centering
   \includegraphics[width=2\columnwidth]{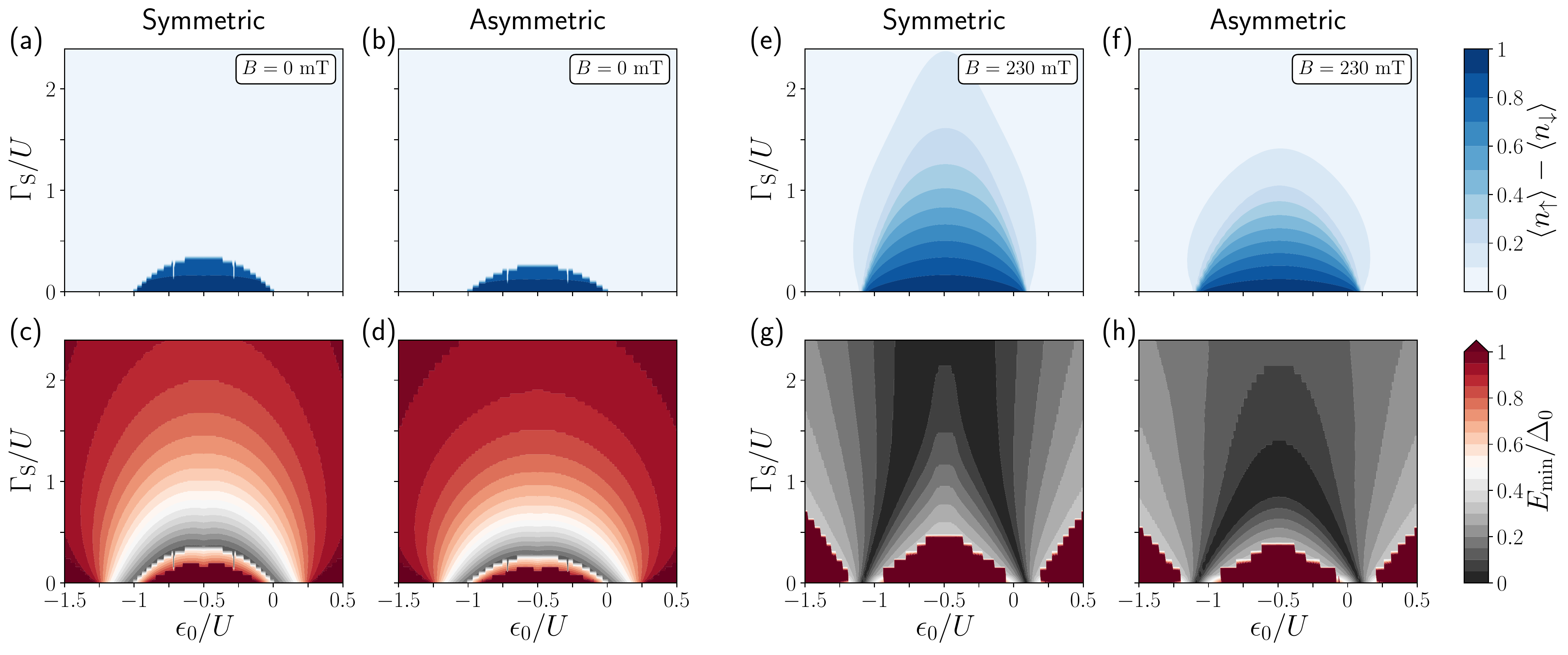}
   \caption{\textbf{QD phase diagrams for $n=0$ and $n=2$.} (a,b) Mean-field dot-level spin polarization $P=\langle n_{\uparrow}\rangle-\langle n_{\downarrow}\rangle$ vs $\Gamma_{\rm S}/U$ and $\epsilon_0/U$ at the center of the $n=0$ LP lobe ($B=0$ mT) in the destructive LP regime: (a) a symmetric coupling ($\bm{r}_0=0$) and (b) an asymmetric one ($|\bm{r}_0|=R/2$). In both cases, we have the usual dome shape for the singlet-doublet transition boundary, where the system has a singlet ground state ($P\approx 0$) in the limit of large coupling ($\Gamma_{\rm S}/U\gtrsim 1$). (c,d) Plot of the minimum excitation energy normalized to the zero-field Al gap. The singlet-doublet boundary is associated to a zero energy parity crossing (black). (e-h) Same but at the center of the $n=2$ LP lobe ($B=230$ mT), in the destructive LP regime as well. Notice that the phase diagram for $n=2$ is very similar to the $n=1$ case, shown in Fig. \ref{fig:3} of the main text.}
   \label{fig:SM1}
\end{figure*}

\begin{table}[htb]
\centering
\caption{Parameters used for the extended Anderson model. We take realistic values extracted from Ref.~\onlinecite{Valentini:S21} .}

\renewcommand{\arraystretch}{1.5}
\newcolumntype{m}{>{\hsize=0.5\hsize}X}
\newcolumntype{s}{>{\hsize=0.82\hsize}X}
\newcolumntype{l}{>{\hsize=1.09\hsize}X}
\newcolumntype{h}{>{\hsize=0.75\hsize}X}
\newcolumntype{j}{>{\hsize=1\hsize}X}

\begin{tabularx}{0.38\textwidth}{ |m|h|h| }
  \multicolumn{3}{c}{Quantum-Dot Hamiltonian } \\
  \hline \hline
  $g=14$ & $U=1$ meV & $m_{\rm D}=0.023m_{0}$ \\
  \hline
\end{tabularx} \ \ \ \ \
\begin{tabularx}{0.15\textwidth}{ |j| }
  \multicolumn{1}{c}{Normal lead} \\
  \hline \hline
  $\Gamma_{\rm N}=10^{-3}$ meV \\
  \hline
\end{tabularx}  \ \ \ \ \
\begin{tabularx}{0.25\textwidth}{ |m|m| }
  \multicolumn{2}{c}{Geometrical parameters} \\
  \hline \hline
  $R=65$ nm & $d=25$ nm \\
  \hline
\end{tabularx} 

\vspace*{0.22 cm}

\begin{tabularx}{0.48\textwidth}{ |s|l|l| }
  \multicolumn{3}{c}{Superconducting Hamiltonian } \\
  \hline \hline
  $\Delta_0=0.2$ meV & $m_{\rm S}=m_0$ & $t_{\theta}=0.01$ meV \\
  \hline
\end{tabularx} \ \ \ \ \
\begin{tabularx}{0.48\textwidth}{ |s|l|l| }
  \multicolumn{3}{c}{Other parameters } \\
  \hline \hline
  $\xi=185$ nm & $T=10$ mK & $\beta=1$ \\
  \hline
\end{tabularx}

\vspace*{0.22 cm}

\label{Table:Params_Anderson}
\end{table}

\subsection{Further analysis}
\label{Subsec:further_analysis}
To gain more insight into this problem, in this section we show some plots not included in the main text. We start by studying the phase diagram for the zeroth-lobe, depicted in Figs. \ref{fig:SM1}(a-d) (the same figure for the first lobe is plotted in Fig. 3 of the main text). In Fig. \ref{fig:SM1}(a), we show the spin polarization $P=\langle n_{\uparrow}\rangle-\langle n_{\downarrow}\rangle$ versus the coupling to the superconductor $\Gamma_{\rm S}$ and the dot level $\epsilon_0$, for the case of a symmetric coupling to the superconductor ($r_0=0$). Blue regions means that the system is spin-polarized, and therefore, it is in a doublet phase; while white regions implies that the system is in a singlet phase. The boundary between both phases exhibits the usual dome-shape profile, whose maximum is at $\Gamma_{\rm S}\simeq U/\pi$ .  Fig. \ref{fig:SM1}(b) shows the asymmetric case ($r_0=R/2$), which in the $n=0$ lobe exhibits a very similar dome-shape. In Figs. \ref{fig:SM1} (c,d), we also show the corresponding minimum excitation energy $E_\textrm{min}$ as in the main text. Black contours indicate the boundary $E_\textrm{min}=0$ for which the quantum phase transition occurs.

\begin{figure*}
   \centering
   \includegraphics[width=2\columnwidth]{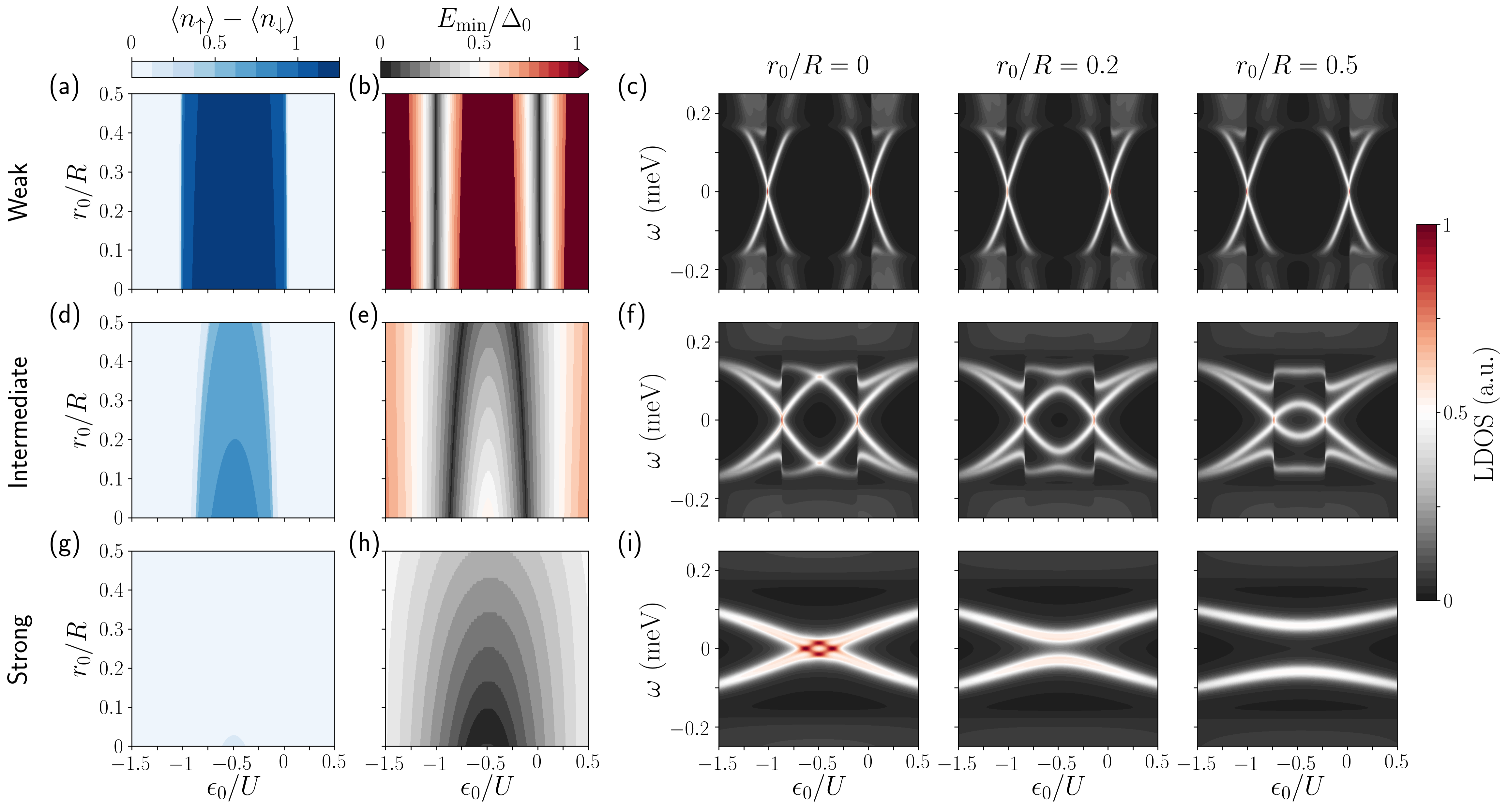}
   \caption{\textbf{QD phase diagram vs QD-S coupling asymmetry $\bf{r_0}$.} Mean-field dot-level spin polarization (a) and minimum excitation energy (b) vs $r_0/R$ and $\epsilon_0/U$ at the center of the $n=1$ lobe ($B=115$ mT) in the destructive LP regime, for a weak coupling to the superconductor ($\Gamma_{\rm S}=0.08$ meV). In (c) we show the LDOS for three different values of $r_0$: the symmetric case $r_0=0$, $r_0/R=0.2$ and $r_0/R=0.5$. In (d-f) and (g-i), we show the same simulations for intermediate ($\Gamma_{\rm S}=0.2$ meV) and a strong ($\Gamma_{\rm S}=2$ meV) coupling to the superconductor, respectively.}
   \label{fig:SM2}
\end{figure*}

For completeness, we show in Figs. \ref{fig:SM1}(e-h) the same simulations but for the second lobe. The spin-polarization phase diagrams of (e) and (f) exhibit a similar chimney and dome like profiles as for the first lobe (see Fig. \ref{fig:3} of the main text). The same occurs for the minimum excitation energy diagram of (g,h), although the energy states are pushed more strongly towards zero energy because the magnitude of the parent gap decreases as we incre ase the lobe number due to the effect of finite-shell thickness in the LP mechanism.

\begin{figure}
   \centering
   \includegraphics[width=1\columnwidth]{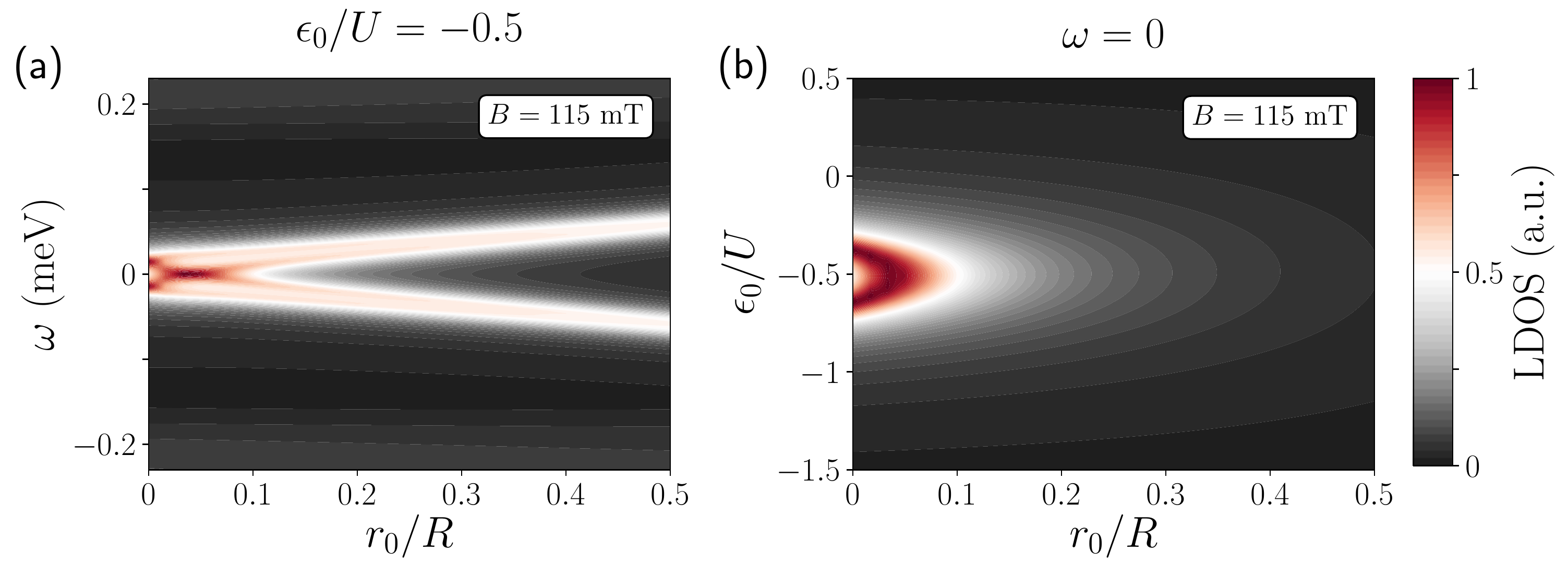}
   \caption{\textbf{Evolution of the near-zero-energy anomaly with QD-S coupling asymmetry $r_0/R$.} LDOS at strong coupling to the superconductor ($\Gamma_{\rm S}=2$ meV) and within the first lobe ($B=115$ mT) vs QD-S coupling asymmetry $r_0/R$ and (a) vs energy $\omega$ at half-filling ($\epsilon_0/U=-0.5$) or (b) vs the QD energy level $\epsilon_0/U$ at $\omega=0$. These figures show that the ABSs remain robustly close to zero energy as long as the degree of QD-S coupling asymmetry is $r_0\lesssim 0.1R$.}
   \label{fig:SM3}
\end{figure}

In Fig. \ref{fig:SM2}, we analyze the behavior of the phase diagram and LDOS versus the dot level $\epsilon_0$, that may be tuned experimentally through the backgate potential, and the QD-S coupling asymmetry $r_0/R$. Recall that, microscopically, the latter represents the wave function mean position inside the covered region with respect to wire's axis, informing of its degree of axial symmetry. The first row [Figs. \ref{fig:SM2}(a-c)] shows the results for the weak coupling regime ($\Gamma_{\rm S}/U\ll 1$), the second row [Figs. \ref{fig:SM2}(d-f)] for an intermediate regime ($\Gamma_{\rm S}/U\lesssim 1$), and the third row [Figs. \ref{fig:SM2}(g-i)] for the strong one ($\Gamma_{\rm S}/U\gg 1$). 

In the weak coupling regime (first row), the parity crossings [black contours in Fig. \ref{fig:SM2}(b)] are influenced only weakly by the QD-S coupling asymmetry. At $r_0=0$ there are two crossing, one at $\epsilon_0/U=-1$ and another one at $\epsilon_0=0$. Between these two crossings, the system is in a doublet phase, as illustrated by the spin polarization phase diagram in (a). As one increases the asymmetry (larger $r_0/R$ value) the phase diagram shows a slight deviation from this picture, with the zero-energy crossings slowly converging towards $\epsilon_0/U=-0.5$. This convergence is due to an effective increase of the coupling to the superconductor as the QD moves closer to one side of the S cylinder. 
Panels (c) in Fig. \ref{fig:SM2} show the evolution of the LDOS in this regime. Together with the weak evolution of the zero-energy crossings mentioned previously, these plots further illustrate the evolution from the ``eyebrow" shape for $r_0=0$ (left), where the upper and lower Andreev bound states (ABSs) touch at $\epsilon_0/U=-0.5$, to the (almost) conventional ``eye"-shape LDOS at $r_0/R=0.5$, where the two ABSs anticross.

In the intermediate coupling regime (second row), the polarization phase diagram [Fig. \ref{fig:SM2}(d)] shows a more dramatic evolution with $r_0/R$, which exhibits a dome-like profile. In fact, this phase diagram shows that one can drive the system through a singlet-doublet transition by just varying the QD-S coupling asymmetry. The minimum excitation-energy phase diagram in (e) also illustrates that the states are closer to zero energy with respect to the weak coupling case, due to the stronger coupling with the superconductor, as explained in the main text. 

\begin{figure*}
   \centering
   \includegraphics[width=1.5\columnwidth]{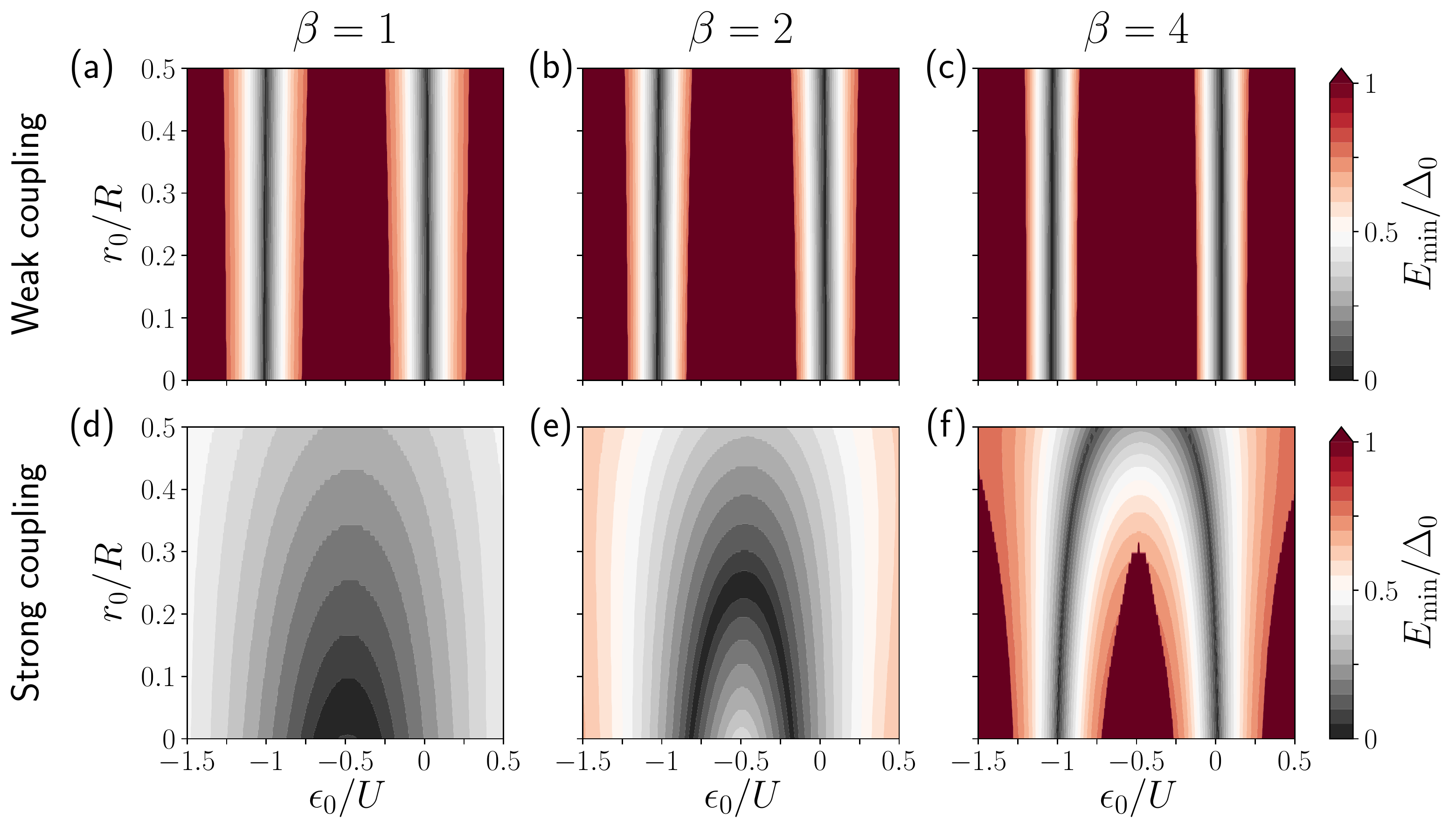}
   \caption{\textbf{QD phase diagram for different QD's inverse widths.} Minimum excitation energy versus $r_0/R$ and $\epsilon_0/U$ at the center of the $n=1$ lobe ($B=115$ mT) in the destructive LP regime, for a weak coupling to the superconductor ($\Gamma_{\rm S}=0.08$ meV), (a-c), and for a strong coupling ($\Gamma_{\rm S}=2$ meV), (d-f).}
   \label{fig:SM4}
\end{figure*}

Finally, in Figs. \ref{fig:SM2}(g,h), we show the phase diagram in the strong coupling regime. In this case, the doublet phase shrinks to a the narrow ``chimney" region in the $\epsilon_0,\Gamma_S$ plane (see Fig. \ref{fig:3} in the main text), but only at small enough $r_0$ and temperature. Thus, the doublet phase becomes essentially invisible in the spin polarization phase diagram of (g). The evolution with $r_0/R$ is better appreciated in the LDOS of Fig. \ref{fig:SM2}(f). 
When $r_0/R\sim 0$ there are zero-energy crossings signaling the singlet-doublet quantum phase transition at the chimney boundaries. As $r_0/R$ grows, the chimney quickly collapses into a dome in the $\epsilon_0,\Gamma_S$ plane, and the LDOS evolves into the regular singlet-phase LDOS of the conventional SIA model at strong coupling.

To better understand the robustness of the near-zero-energy modes depicted in Fig. \ref{fig:SM2}(i), we show in Fig. \ref{fig:SM3}(a) the LDOS evolution at half-filling, i.e., at $\epsilon_0/U=-0.5$, versus the QD-S coupling asymmetry $r_0/R$. In Fig. \ref{fig:SM3}(b) we show the same quantity versus the QD energy level $\epsilon_0$ at $\omega=0$. For $r_0\lesssim 0.1R$, these ABSs remain  close to zero energy, in such a way that if temperature is large enough, they are indistinguishable from a single zero-energy peak. As we discussed in Appendix \ref{AppA}, this may be a rather realistic scenario that could be relevant when interpreting the measurements in experiments \cite{Vaitiekenas:S20,Valentini:S21}. Conversely, for $r_0>0.1R$ the near-zero-energy modes split and disperse linearly with the dot position $r_0$. As explained in the main text, this phenomenon occurs when the coupling between the QD and the S is no longer axially symmetric, and therefore, non-zero superconducting pairing correlations are induced into the QD. These in turn create an induced gap in the QD that, as shown here, is noticeable for $r_0>0.1R$ (at $\Gamma_{\rm S}=2$ meV).

To complement our study, we analyze in Fig. \ref{fig:SM4} the behavior of the phase diagram versus the parameter $\beta$ of Eq. \eqref{Eq:t(theta)}. This parameter, as introduced in Sec. \ref{Sec:QD-S_coupling}, models the QD's inverse width. Assuming a Gaussian-like profile for the QD wavefunction, this parameter is proportional to the inverse of standard deviation $\sigma$ of the QD's wavefunction (i.e., $\beta\sim 1/\sigma^2$), which we actually compute in our microscopic simulations [see blue curve in Fig. \ref{fig:SM0}(c)]. This means that the larger $\beta$, the more localized the QD wavefunction is. The plots in Fig. \ref{fig:SM4} show the minimum excitation-energy diagram as a function of the QD-S coupling asymmetry $r_0/R$ and its energy $\epsilon_0/U$ for different cases: $\beta=1$ in (a), $\beta=2$ in (b) and $\beta=4$ in (c), in the weak coupling regime. In (d-f) we show the same simulations in the strong coupling one. In the former, we observe no change in the position of the zero-energy crossing as $\beta$ is increased. The reason is that the QD is weakly coupled to the superconductor and therefore a change in the exponential of Eq. \eqref{Eq:t(theta)} makes almost no difference. However, for $\beta=4$ in (c) the states disperse more dramatically with $\epsilon_0$, as revealed by a larger dark red region, meaning that the states are above the gap there. This is because for this large $\beta$ the coupling is highly suppressed, and therefore we recover the atomic limit for the QD states. On the other hand, there is a remarkable change with $\beta$ in the strong coupling regime [Figs. \ref{fig:SM4}(d-f)]. As explained before, a larger $\beta$ implies a more point like QD, and therefore a reduced overlap and coupling to the superconductor. Hence, the profile for a large $\beta$ in the strong coupling case (f) is similar to those in the weak coupling (a-c). We thus conclude that $\beta$ and $\Gamma_{\rm S}$ play a similar role in our model.

\section{Kondo correlation effects}
\label{AppC}

All results in the main text and in this Appendix were derived within a mean-field treatment of QD interactions. This approximation is only strictly valid at small $\Gamma_{\rm S}/U$, for which the Kondo temperature $T_{\rm K}$ is smaller than $\Delta$. At stronger couplings, correlation effects have to be included in the treatment of $\Sigma^{\rm U}_\sigma$, see Eq. (2) in the main text. Their effect has been studied in the literature of S-QD-S junctions using a variety of non-perturbative techniques~\cite{Oguri:JPSJ04, Meng:PRB09, Martin-Rodero:AIP11, Zonda:SR15, Zitko:PRB15, Wentzell:PRB16, lee:PRB17, Meden:JOP19}. It has been shown in particular that, at junctions with phase $\phi = \pi$, the chimney like phase boundary survives correlations, and actually grows in width relative to the mean field result~\cite{Zonda:SR15}. The doublet ground state at strong coupling evolves in this case into an overscreened two-channel-Kondo doublet. We speculate that this is also the case in the present problem, although a rigorous treatment would require a non-perturbative calculation within a multichannel Kondo context~\cite{Muhlschlegel:ZFPA68,Nozieres:JDP80,Hewson:93}, where the channels are the occupied modes in the shell.
Further Kondo correlations can be induced by the N probe \cite{Zitko:PRB15} on doublet ground states. Much like in $\phi = \pi$ S-QD-S junctions \cite{Zalom:PRB21}, a finite $\Gamma_N$ may result in a zero-energy Kondo peak at sufficiently low temperatures, in addition to the near-zero energy features discussed here.

\bibliography{biblio}

\end{document}